# Dissipative Stabilization of Linear Systems with Time-Varying General Distributed Delays (Complete Version)[⋆]

Qian Feng[a,∗], Sing Kiong Nguang[b], Wilfrid Perruquetti[a]

[a]*École Centrale de Lille, CRIStAL, UMR CNRS 9189 Cite Scientifique, 59651, Villneuve d'ascq Cedex, France*
[b]*Department of Electrical and Computer Engineering, The University of Auckland, Auckland 1010, New Zealand*

**Abstract**

New methods are developed for the stabilization of a linear system with general time-varying distributed delays existing at the system's states, inputs and outputs. In contrast to most existing literature where the function of time-varying delay is continuous and bounded, we assume it to be bounded and measurable. Furthermore, the distributed delay kernels can be any square-integrable function over a bounded interval, where the kernels are handled directly by using a decomposition scenario without using approximations. By constructing a Krasovski functional via the application of a novel integral inequality, sufficient conditions for the existence of a dissipative state feedback controller are derived in terms of matrix inequalities without utilizing the existing reciprocally convex combination lemmas. The proposed synthesis (stability) conditions, which take dissipativity into account, can be either solved directly by a standard numerical solver of semidefinite programming if they are convex, or reshaped into linear matrix inequalities, or solved via a proposed iterative algorithm. To the best of our knowledge, no existing methods can handle the synthesis problem investigated in this paper. Finally, numerical examples are presented to demonstrate the effectiveness of the proposed methodologies.

## 1. Introduction

Time delays exist in systems affected by transportation and aftereffects Richard (2003). For certain real-time application such as the models in Anthonis *et al.* (2007); Molnár & Insperger (2015), delays can be time-varying. It is of great research interest to investigate a system with bounded time-varying delays since it can be applied in modeling sampled-data Fridman *et al.* (2004) or networked control systems (NCSs) Hespanha *et al.* (2007). One can find many existing results in the literature pertaining to the stability analysis Jiang (2006); Seuret *et al.* (2013); Kwon *et al.* (2016); Qian *et al.* (2018) and stabilization Jiang (2005); Fridman (2006); Mohajerpoor *et al.* (2018) of linear time-varying delay systems with a bounded continuous time-varying delay. The methods in the aforementioned references are based on the construction of Krasovski functionals (KF) Briat (2014), where the time-varying delay is embedded only in $\boldsymbol{x}(t-r(t))$. It has been

[⋆]This work was partially supported by the ANR (Agence nationale de la recherche) Project Finite4SoS (ANR-15-CE23-0007).

[∗]Corresponding author
*Email addresses:* qfen204@aucklanduni.ac.nz, qian.feng@centralelille.fr (Qian Feng), sk.nguang@auckland.ac.nz (Sing Kiong Nguang), wilfrid.perruquetti@centralelille.fr (Wilfrid Perruquetti)

shown in Gao *et al.* (2008, 2010) that the KF method for linear systems with $\boldsymbol{x}(t-r(t))$ can be utilized to solve synthesis problems of NCSs. It is worthy mention that unlike systems with constant delays, frequency-domain-based approaches Breda *et al.* (2005); Michiels & Niculescu (2014); Gehring *et al.* (2014); Vyhlídal & Zítek (2014) may not be usable to analyze the spectrum of a system if the exact expression of $r(t)$ is unknown.

It has been pointed out in Goebel *et al.* (2011) that the digital communication channel of NCSs with stochastic packet delays and loss can be modeled by distributed delays. Moreover, the results in Yan *et al.* (2019) have shown that a networked control system with a network channel delay stabilized by an event-triggered $\mathcal{H}^{\infty}$ controller can be modeled as a distributed delay system, where the delay is of constant values. To the best of our knowledge however, no existing methods can handle the stabilization problem of systems with general distributed delays where the delay function is time-varying and unknown. In Theorem 2 of Zhou *et al.* (2012), a method of stabilizing systems in the form of $\dot{\boldsymbol{x}}(t) = A\boldsymbol{x}(t) + \int_{-r(t)}^{0} B(\tau)\boldsymbol{u}(t+\tau)\mathrm{d}\tau$ is proposed. Nevertheless, all the poles of $A$ in Zhou *et al.* (2012) are assumed to be located on the imaginary axis, and the lower bound of $r(t)$ is restricted to $0 < r(t) \leq r_2$. The stability of positive linear systems with distributed time-varying delays is investigated in Ngoc (2013); Cui *et al.* (2018). Although the method in Ngoc (2013) does include criteria to determine the stability of non-positive linear systems, the structure of the delay function $r(\cdot) \in \mathbb{C}\left(\mathbb{R}_{\geq 0} \fatsemi (0, r_2]\right)$ therein is still restrictive. On the other hand, the synthesis (stability analysis) methods proposed in Münz *et al.* (2009); Goebel *et al.* (2011); Gouaisbaut *et al.* (2015); Seuret *et al.* (2015); Feng & Nguang (2016), which are developed to handle linear distributed delay systems with constant delay values, may not be easily extended to cope with systems with an unknown time-varying delay. This is especially true for the approximation approaches in Münz *et al.* (2009); Goebel *et al.* (2011); Gouaisbaut *et al.* (2015); Seuret *et al.* (2015), since the approximation coefficients can become nonlinear with respect to $r(t)$ if the distributed delay kernels are approximated over $[-r(t), 0]$. Consequently, it is obvious that new methods should be developed for the stabilization (stability analysis) of linear systems with general distributed time-varying delays.

In this paper, new methods for the stabilization of a linear system with general distributed time-varying delays are developed based on the construction of a general Krasovski functional, where the time-varying delay function $r(\cdot)$ is unknown but measurable and bounded by given values $0 \leq r_1 \leq r_2$, $r_2 > 0$. Our system contains distributed delays at the system's states, inputs and outputs, where the delay kernels can be any $\mathbb{L}^2$ function over a given interval. To ensure that the proposed methods are denoted by linear matrix inequalities (LMIs) with finite-dimensions, a novel integral inequality is proposed where the symmetric matrix of the inequality's lower bound is not a function of $r(t)$ but $r_1$ and $r_2$. By using this inequality in constructing a general KF, sufficient conditions for the existence of a state feedback controller, which ensure that the system is stable and dissipative with a supply function, are derived in terms of matrix inequalities summarized in the first theorem of this paper. The synthesis condition of the first theorem has a bilinear matrix inequality (BMI) if a stabilization problem is considered, whereas it becomes convex if non-stabilization scenarios are concerned. To circumvent the difficulties of solving non-convex conditions, a second theorem denoted by LMIs is proposed via the application of Projection Lemma Gahinet & Apkarian (1994). Next, an iterative algorithm, based on the inner convex approximation scheme developed in Dinh *et al.* (2012), is proposed for solving the BMI in the first theorem, where the algorithm can be initiated by a feasible solution of the second theorem. To the best of our knowledge, no existing methods in the peer-reviewed literature can handle the synthesis problem in this paper. Finally, two numerical examples are presented to demonstrate the effectiveness of our proposed methodologies.

The major contributions of this paper are summarized as follows:

- We believe the dissipative synthesis problem investigated in this paper cannot be dealt with by any existing method. Our model is sufficiently general with respect to generality of $r(t)$ and distributed delay kernels. A bounded and measurable delay function can be particularly useful to model discontinuous time-varying delays. Finally, the proposed methods only require the use of standard algorithms for semidefinite programming (SDP) without asking for nonlinear solvers.

- The handling of distributed delay kernels in this paper, which is based on the application of a decomposition approach, allows one to consider any $\mathbb{L}^2$ function over an interval, even the interval is related to $r(t)$. This avoids the use of any form of approximations so that no nonlinear terms of $r(t)$ will be introduced into the proposed synthesis conditions.

- The proposed integral inequality allows one to construct Krasovski functionals without utilizing the reciprocally-convex-combination type of lemmas Park *et al.* (2011); Seuret & Gouaisbaut (2017); Zhang *et al.* (2017a,b); Seuret *et al.* (2018) which may not be capable of providing tractable solutions to the problem considered in this paper.

The rest of the paper is outlined as follows. The synthesis problem investigated in this paper is first formulated in Section 2 where we explain the principle of the decomposition approach. Next, the main results on dissipative stabilization are presented in Section 3, which are summarized in Theorem 1 and 2 and Algorithm 1. Finally, numerical examples and their simulation results are presented in Section 4 prior to the final conclusion.

**Notation**

Let $\mathcal{Y}^{\mathcal{X}} = \{f(\cdot) : f(\cdot) \text{ is a function from } \mathcal{X} \text{ onto } \mathcal{Y}\}$ and $\mathbb{R}_{\geq a} = \{x \in \mathbb{R} : x \geq a\}$ and $\mathbb{S}^n = \{X \in \mathbb{R}^{n\times n} : X = X^\top\}$. $\mathbb{C}(\mathcal{X} ; \mathbb{R}^n) := \left\{\boldsymbol{f}(\cdot) \in (\mathbb{R}^n)^{\mathcal{X}} : \boldsymbol{f}(\cdot) \text{ is continuous on } \mathcal{X}\right\}$ and $\mathbb{C}^k([a,b] ; \mathbb{R}^n) := \left\{\boldsymbol{f}(\cdot) \in \mathbb{C}([a,b] ; \mathbb{R}^n) : \frac{\mathrm{d}^k \boldsymbol{f}(x)}{\mathrm{d}x^k} \in \mathbb{C}([a,b] ; \mathbb{R}^n)\right\}$ where the derivatives at $a$ and $b$ are one-sided. Moreover, $\mathbb{M}_{\boldsymbol{\mathcal{L}}(\mathcal{X})/\boldsymbol{\mathcal{B}}(\mathbb{R})}(\mathcal{X} ; \mathbb{R}) := \left\{f(\cdot) \in \mathcal{X}^{\mathbb{R}} : \forall \mathcal{Y} \in \boldsymbol{\mathcal{B}}(\mathbb{R}),\ f^{-1}(\mathcal{Y}) \in \boldsymbol{\mathcal{L}}(\mathcal{X})\right\}$ denotes the space of all $\boldsymbol{\mathcal{L}}(\mathcal{X})/\boldsymbol{\mathcal{B}}(\mathbb{R})$ measurable functions from $\mathcal{X}$ onto $\mathbb{R}$, where $\boldsymbol{\mathcal{L}}(\mathcal{X})$ contains all the subsets of $\mathcal{X}$ which are Lebesgue measurable, and $\boldsymbol{\mathcal{B}}(\mathbb{R})$ is the Borel $\sigma$–algebra on $\mathbb{R}$. Note that we frequently abbreviate $\mathbb{M}_{\boldsymbol{\mathcal{L}}(\mathcal{X})/\boldsymbol{\mathcal{B}}(\mathbb{R})}(\mathcal{X} ; \mathbb{R})$ as $\mathbb{M}(\mathcal{X} ; \mathbb{R})$ when the context is clear. In addition, we define $\mathbb{L}^p(\mathcal{X} ; \mathbb{R}^n) = \left\{\boldsymbol{f}(\cdot) \in \mathbb{M}_{\boldsymbol{\mathcal{L}}(\mathcal{X})/\boldsymbol{\mathcal{B}}(\mathbb{R}^n)}(\mathcal{X} ; \mathbb{R}^n) : \left\|\boldsymbol{f}(\cdot)\right\|_p < +\infty\right\}$ with the semi-norm $\left\|\boldsymbol{f}(\cdot)\right\|_p := \left(\int_{\mathcal{X}} \left\|\boldsymbol{f}(x)\right\|_2^p \mathrm{d}x\right)^{\frac{1}{p}}$ where $\mathcal{X} \subseteq \mathbb{R}^n$. Moreover, we use $\widetilde{\forall} x \in \mathcal{X}$ to denote the meaning of for almost all $x \in \mathcal{X}$ with respect to the Lebesgue measure. $\mathbf{Sy}(X) := X + X^\top$ stands for the sum of a matrix with its transpose. We use $\mathbf{Col}_{i=1}^n x_i := \left[\mathbf{Row}_{i=1}^n x_i^\top\right]^\top = \left[x_1^\top \cdots x_i^\top \cdots x_n^\top\right]^\top$ to denote a column vector containing a sequence of mathematical objects (scalars, vectors, matrices etc.). The symbol $*$ is used to indicate $[*]YX = X^\top YX$ or $X^\top Y[*] = X^\top YX$ or $\left[\begin{smallmatrix} A & B \\ * & C \end{smallmatrix}\right] = \left[\begin{smallmatrix} A & B \\ B^\top & C \end{smallmatrix}\right]$. $\mathsf{O}_{n,m}$ denotes a $n \times m$ zero matrix which can be abbreviated by $\mathsf{O}_n$ if $n = m$, while $\mathbf{0}_n$ represents a $n \times 1$ column vector. We frequently use $X \oplus Y = \left[\begin{smallmatrix} X & \mathsf{O} \\ * & Y \end{smallmatrix}\right]$ to denote the diagonal sum of two matrices. $\otimes$ stands for the Kronecker product. The order of matrix operations is *matrix (scalars) multiplications* $> \otimes > \oplus > +$ throughout the paper. Finally, empty matrices $[\ ]_{0,m}$, $[\ ]_{m,0}$, $[\ ]_{0,0}$ (Stoer & Witzgall, 1970, See I.7) with $m \in \mathbb{N}$, which follow the same definition and programming rules in Matlab©, are applied in this paper to render our synthesis conditions capable of handling the case of $r_1 = 0; r_2 > 0$ or $r_1 = r_2$. Note that we define $\mathbf{Col}_{i=1}^{n} x_i = [\,]_{0,m}$ if $n < 1$, where $[\,]_{0,m}$ is an empty matrix with an appropriate column dimension $m \in \mathbb{N}$ based on specific contexts.

## 2. Problem formulation

Consider a linear distributed delay system

$$
\begin{aligned}
&\dot{\boldsymbol{x}}(t) = A_1\boldsymbol{x}(t) + \int_{-r(t)}^{0} \widetilde{A}_2(\tau)\boldsymbol{x}(t+\tau)\mathsf{d}\tau + B_1\boldsymbol{u}(t) + \int_{-r(t)}^{0} \widetilde{B}_2(\tau)\boldsymbol{u}(t+\tau)\mathsf{d}\tau + D_1\boldsymbol{w}(t), \quad \forall t \geq t_0 \\
&\boldsymbol{z}(t) = C_1\boldsymbol{x}(t) + \int_{-r(t)}^{0} \widetilde{C}_2(\tau)\boldsymbol{x}(t+\tau)\mathsf{d}\tau + B_4\boldsymbol{u}(t) + \int_{-r(t)}^{0} \widetilde{B}_5(\tau)\boldsymbol{u}(t+\tau)\mathsf{d}\tau + D_2\boldsymbol{w}(t), \\
&\forall \theta \in [-r_2, 0], \quad \boldsymbol{x}(t_0+\theta) = \boldsymbol{\phi}(\tau), \quad r(\cdot) \in \mathbb{M}\left(\mathbb{R} \,;\, [r_1, r_2]\right)
\end{aligned} \tag{1}
$$

to be stabilized, where $t_0 \in \mathbb{R}$ and $\boldsymbol{\phi}(\cdot) \in \mathbb{C}([-r_2, 0] \,;\, \mathbb{R}^n)$, and $r_2 > 0$, $r_2 \geq r_1 \geq 0$ are given constants. Furthermore, $\boldsymbol{x} : [t_0 - r_2, \infty) \to \mathbb{R}^n$ satisfies (1), $\boldsymbol{u}(t) \in \mathbb{R}^p$ denotes input signals, $\boldsymbol{w}(t) \in \mathbb{L}^2\left([t_0, +\infty) \,;\, \mathbb{R}^q\right)$ represents disturbance, and $\boldsymbol{z}(t) \in \mathbb{R}^m$ is the regulated output. The size of the given state space parameters in (1) is determined by the values of $n \in \mathbb{N}$ and $m; p; q \in \mathbb{N}_0 := \mathbb{N} \cup \{0\}$. Finally, the matrix-valued distributed delays in (1) satisfy

$$
\begin{aligned}
&\widetilde{A}_2(\cdot) \in \mathbb{L}^2([-r_2, 0] \,;\, \mathbb{R}^{n \times n}), \quad \widetilde{C}_2(\cdot) \in \mathbb{L}^2([-r_2, 0] \,;\, \mathbb{R}^{m \times n}) \\
&\widetilde{B}_2(\cdot) \in \mathbb{L}^2([-r_2, 0] \,;\, \mathbb{R}^{n \times p}), \quad \widetilde{B}_5(\cdot) \in \mathbb{L}^2([-r_2, 0] \,;\, \mathbb{R}^{m \times p}).
\end{aligned} \tag{2}
$$

**Remark 1.** Systems with distributed delays and a time-varying delay function can be found among the models of neural networks Ge *et al.* (2019); Dong *et al.* (2019).

The distributed delays in (2) are infinite-dimensional. In order to construct synthesis constraints with finite dimensions for (1), we propose a decomposition scenario as follows.

**Proposition 1.** *The conditions in* (2) *holds if and only if there exist* $\boldsymbol{f}_1(\cdot) \in \mathbb{C}^1([-r_2, 0] \,;\, \mathbb{R}^{d_1})$, $\boldsymbol{f}_2(\cdot) \in \mathbb{C}^1([-r_2, 0] \,;\, \mathbb{R}^{d_2})$, $\boldsymbol{\varphi}_1(\cdot) \in \mathbb{L}^2([-r_2, 0] \,;\, \mathbb{R}^{\delta_1})$, $\boldsymbol{\varphi}_2(\cdot) \in \mathbb{L}^2([-r_2, 0] \,;\, \mathbb{R}^{\delta_2})$, $M_1 \in \mathbb{R}^{d_1 \times \kappa_1}$, $M_2 \in \mathbb{R}^{d_2 \times \kappa_2}$, $A_2 \in \mathbb{R}^{n \times \kappa_1 n}$, $A_3 \in \mathbb{R}^{n \times \kappa_2 n}$, $B_2 \in \mathbb{R}^{n \times \kappa_1 p}$, $B_3 \in \mathbb{R}^{n \times \kappa_2 p}$, $C_2 \in \mathbb{R}^{m \times \kappa_1 n}$, $C_3 \in \mathbb{R}^{m \times \kappa_2 n}$, $B_5 \in \mathbb{R}^{m \times \kappa_1 p}$ *and* $B_6 \in \mathbb{R}^{m \times \kappa_2 p}$ *such that*

$$
\forall \tau \in [-r_1, 0], \quad \widetilde{A}_2(\tau) = A_2\left(\widehat{\boldsymbol{f}}_1(\tau) \otimes I_n\right), \quad \widetilde{B}_2(\tau) = B_2\left(\widehat{\boldsymbol{f}}_1(\tau) \otimes I_p\right), \tag{3}
$$

$$
\forall \tau \in [-r_2, -r_1], \quad \widetilde{A}_2(\tau) = A_3\left(\widehat{\boldsymbol{f}}_2(\tau) \otimes I_n\right), \quad \widetilde{B}_2(\tau) = B_3\left(\widehat{\boldsymbol{f}}_2(\tau) \otimes I_p\right), \tag{4}
$$

$$
\forall \tau \in [-r_1, 0], \quad \widetilde{C}_2(\tau) = C_2\left(\widehat{\boldsymbol{f}}_1(\tau) \otimes I_n\right), \quad \widetilde{B}_5(\tau) = B_5\left(\widehat{\boldsymbol{f}}_1(\tau) \otimes I_p\right), \tag{5}
$$

$$
\forall \tau \in [-r_2, -r_1], \quad \widetilde{C}_2(\tau) = C_3\left(\widehat{\boldsymbol{f}}_2(\tau) \otimes I_n\right), \quad \widetilde{B}_5(\tau) = B_6\left(\widehat{\boldsymbol{f}}_2(\tau) \otimes I_p\right), \tag{6}
$$

$$
\forall \tau \in [-r_2, 0], \quad \frac{\mathsf{d}\boldsymbol{f}_1(\tau)}{\mathsf{d}\tau} = M_1\widehat{\boldsymbol{f}}_1(\tau), \quad \frac{\mathsf{d}\boldsymbol{f}_2(\tau)}{\mathsf{d}\tau} = M_2\widehat{\boldsymbol{f}}_2(\tau) \tag{7}
$$

$$
\mathsf{G}_1 = [\,]_{0 \times 0} \ \text{ or } \ \mathsf{G}_1 \succ 0, \quad \mathsf{G}_1 := \int_{-r_1}^{0} \widehat{\boldsymbol{f}}_1(\tau)\widehat{\boldsymbol{f}}_1^{\top}(\tau)\mathsf{d}\tau \tag{8}
$$

$$
\mathsf{G}_2 = [\,]_{0 \times 0} \ \text{ or } \ \mathsf{G}_2 \succ 0, \quad \mathsf{G}_2 := \int_{-r_2}^{-r_1} \widehat{\boldsymbol{f}}_2(\tau)\widehat{\boldsymbol{f}}_2^{\top}(\tau)\mathsf{d}\tau \tag{9}
$$

*where* $\kappa_1 = d_1 + \delta_1$, $\kappa_2 = d_2 + \delta_2$ *with* $d_1; d_2; \delta_1; \delta_2 \in \mathbb{N}_0$ *satisfying* $d_1 + d_2 > 0$, *and*

$$\widehat{\boldsymbol{f}}_1(\tau) = \begin{bmatrix} \boldsymbol{\varphi}_1(\tau) \\ \boldsymbol{f}_1(\tau) \end{bmatrix}, \quad \widehat{\boldsymbol{f}}_2(\tau) = \begin{bmatrix} \boldsymbol{\varphi}_2(\tau) \\ \boldsymbol{f}_2(\tau) \end{bmatrix}. \tag{10}$$

*Finally, the derivatives in* (7) *at* $\tau = 0$ *and* $\tau = -r_2$ *are one-sided derivatives. Note that if matrix multiplications in* (3)–(10) *involve any empty matrix, then it follows the definition and properties of empty matrices in Matlab.*

*Proof.* First of all, it is straightforward to see that (2) is implied by (3)–(10) since $\boldsymbol{\varphi}_1(\cdot) \in \mathbb{L}^2([-r_2, 0] ; \mathbb{R}^{\delta_1})$, $\boldsymbol{\varphi}_2(\cdot) \in \mathbb{L}^2([-r_2, 0] ; \mathbb{R}^{\delta_2})$, $\boldsymbol{f}_1(\cdot) \in \mathbb{C}^1([-r_2, 0] ; \mathbb{R}^{d_1}) \subset \mathbb{L}^2([-r_2, 0] ; \mathbb{R}^{d_1})$ and $\boldsymbol{f}_2(\cdot) \in \mathbb{C}^1([-r_2, 0] ; \mathbb{R}^{d_2}) \subset \mathbb{L}^2([-r_2, 0] ; \mathbb{R}^{d_2})$.

Now we start to prove (2) implies the existence of the parameters in Proposition 1 satisfying (3)–(9). Given any $\boldsymbol{f}_1(\cdot) \in \mathbb{C}^1([-r_2, 0] ; \mathbb{R}^{d_1})$, $\boldsymbol{f}_2(\cdot) \in \mathbb{C}^1([-r_2, 0] ; \mathbb{R}^{d_2})$, one can always find appropriate $\boldsymbol{\varphi}_1(\cdot) \in \mathbb{L}^2([-r_2, 0] ; \mathbb{R}^{\delta_1})$, $\boldsymbol{\varphi}_2(\cdot) \in \mathbb{L}^2([-r_2, 0] ; \mathbb{R}^{\delta_2})$ with $M_1 \in \mathbb{R}^{d_1 \times \kappa_1}$ and $M_2 \in \mathbb{R}^{d_2 \times \kappa_2}$ such that (7)–(9) are satisfied with (10), where $\mathsf{G}_1 \succ 0$ and $\mathsf{G}_2 \succ 0$ in (8) infers that the functions in $\widehat{\boldsymbol{f}}_1(\cdot)$ and $\widehat{\boldsymbol{f}}_2(\cdot)$ in (10) are linearly independent[1] in a Lebesgue sense over $[-r_2, 0]$ and $[-r_2, -r_1]$, respectively. This is true since $\frac{\mathsf{d}\boldsymbol{f}_1(\tau)}{\mathsf{d}\tau}(\cdot) \in \mathbb{L}^2([-r_1, 0] ; \mathbb{R}^{d_1})$ and $\frac{\mathsf{d}\boldsymbol{f}_1(\tau)}{\mathsf{d}\tau}(\cdot) \in \mathbb{L}^2([-r_2, -r_2] ; \mathbb{R}^{d_2})$, and the dimensions of $\boldsymbol{\varphi}_1(\tau)$ and $\boldsymbol{\varphi}_2(\tau)$ can be arbitrarily enlarged with more linearly independent functions. Note that if any vector-valued function $\boldsymbol{f}_1(\tau)$, $\boldsymbol{f}_2(\tau)$, $\boldsymbol{\varphi}_1(\tau)$ and $\boldsymbol{\varphi}_2(\tau)$ is $[\,]_{0\times 1}$, then it can be handled by the application of empty matrices as reflected in (8) and (9).

Given any $\boldsymbol{f}_1(\cdot) \in \mathbb{C}^1([-r_2, 0] ; \mathbb{R}^{d_1})$, $\boldsymbol{f}_2(\cdot) \in \mathbb{C}^1([-r_2, 0] ; \mathbb{R}^{d_2})$, we have shown that one can always construct appropriate $\boldsymbol{\varphi}_1(\cdot) \in \mathbb{L}^2([-r_2, 0] ; \mathbb{R}^{\delta_1})$, $\boldsymbol{\varphi}_2(\cdot) \in \mathbb{L}^2([-r_2, 0] ; \mathbb{R}^{\delta_2})$ with $M_1$ and $M_2$ such that the conditions in (7)–(9) are satisfied with (10). As a result, based on the definition of matrix-valued functions and the fact that the dimensions of $\boldsymbol{\varphi}_1(\tau)$ and $\boldsymbol{\varphi}_2(\tau)$ can be arbitrarily increased, one can always construct appropriate constant matrices $A_{2,i}$, $A_{3,i}$, $C_{2,i}$, $C_{3,i}$, $B_{2,i}$, $B_{3,i}$, $B_{5,i}$, $B_{6,i}$ and $\boldsymbol{f}_1(\tau)$, $\boldsymbol{f}_2(\tau)$, $\boldsymbol{\varphi}_1(\tau)$ and $\boldsymbol{\varphi}_2(\tau)$ such that

$$\forall \tau \in [-r_1, 0],\ \widetilde{A}_2(\tau) = \sum_{i=1}^{\kappa_1} A_{2,i} g_i(\tau),\ \ \widetilde{C}_2(\tau) = \sum_{i=1}^{\kappa_1} C_{2,i} g_i(\tau), \tag{11}$$

$$\forall \tau \in [-r_1, 0],\ \widetilde{B}_2(\tau) = \sum_{i=1}^{\kappa_1} B_{2,i} g_i(\tau),\ \ \widetilde{B}_5(\tau) = \sum_{i=1}^{\kappa_1} B_{5,i} g_i(\tau) \tag{12}$$

$$\forall \tau \in [-r_2, -r_1],\ \widetilde{A}_2(\tau) = \sum_{i=1}^{\kappa_2} A_{3,i} h_i(\tau),\ \ \widetilde{C}_2(\tau) = \sum_{i=1}^{\kappa_2} C_{3,i} h_i(\tau), \tag{13}$$

$$\forall \tau \in [-r_2, -r_1],\ \widetilde{B}_2(\tau) = \sum_{i=1}^{\kappa_2} B_{3,i} h_i(\tau),\ \ \widetilde{B}_5(\tau) = \sum_{i=1}^{\kappa_2} B_{6,i} h_i(\tau) \tag{14}$$

$$\boldsymbol{g}^\top(\tau) = \widehat{\boldsymbol{f}}_1^\top(\tau) = \begin{bmatrix} \boldsymbol{\varphi}_1^\top(\tau) & \boldsymbol{f}_1^\top(\tau) \end{bmatrix}^\top, \quad \boldsymbol{h}(\tau) = \widehat{\boldsymbol{f}}_2^\top(\tau) = \begin{bmatrix} \boldsymbol{\varphi}_2^\top(\tau) & \boldsymbol{f}_2^\top(\tau) \end{bmatrix}^\top \tag{15}$$

with $\kappa_1; \kappa_2 \in \mathbb{N}_0$, where $\boldsymbol{f}_1(\tau)$, $\boldsymbol{f}_2(\tau)$, $\boldsymbol{\varphi}_1(\tau)$ and $\boldsymbol{\varphi}_2(\tau)$ satisfy (7)–(9) for some $M_1$ and $M_2$. Now (11)–(14) can be further rewritten as

$$\begin{aligned}
&\forall \tau \in [-r_1, 0],\ \widetilde{A}_2(\tau) = \left[\mathop{\mathsf{Row}}_{i=1}^{\kappa_1} A_{2,i}\right] \left(\widehat{\boldsymbol{f}}_1(\tau) \otimes I_n\right), && \widetilde{C}_2(\tau) = \left[\mathop{\mathsf{Row}}_{i=1}^{\kappa_1} C_{2,i}\right] \left(\widehat{\boldsymbol{f}}_1(\tau) \otimes I_n\right) \\
&\forall \tau \in [-r_2, -r_1],\ \widetilde{A}_2(\tau) = \left[\mathop{\mathsf{Row}}_{i=1}^{\kappa_2} A_{3,i}\right] \left(\widehat{\boldsymbol{f}}_2(\tau) \otimes I_n\right), && \widetilde{C}_2(\tau) = \left[\mathop{\mathsf{Row}}_{i=1}^{\kappa_2} C_{3,i}\right] \left(\widehat{\boldsymbol{f}}_2(\tau) \otimes I_n\right) \\
&\forall \tau \in [-r_1, 0],\ \widetilde{B}_2(\tau) = \left[\mathop{\mathsf{Row}}_{i=1}^{\kappa_1} B_{2,i}\right] \left(\widehat{\boldsymbol{f}}_1(\tau) \otimes I_p\right), && \widetilde{B}_5(\tau) = \left[\mathop{\mathsf{Row}}_{i=1}^{\kappa_1} B_{5,i}\right] \left(\widehat{\boldsymbol{f}}_1(\tau) \otimes I_p\right) \\
&\forall \tau \in [-r_2, -r_1],\ \widetilde{B}_2(\tau) = \left[\mathop{\mathsf{Row}}_{i=1}^{\kappa_2} B_{3,i}\right] \left(\widehat{\boldsymbol{f}}_2(\tau) \otimes I_p\right), && \widetilde{B}_5(\tau) = \left[\mathop{\mathsf{Row}}_{i=1}^{\kappa_2} B_{6,i}\right] \left(\widehat{\boldsymbol{f}}_2(\tau) \otimes I_p\right).
\end{aligned} \tag{16}$$

[1] See Theorem 7.2.10 in Horn & Johnson (2012) for more information

which are in line with the forms in (3)–(6). Given all the aforementioned statements we have presented, then Proposition 1 is proved. ■

**Remark 2.** Proposition 1 provides an effective way to handle the distributed delays in (1). It uses a group of basis functions to decompose the distributed delays without appealing to the application of approximations. The potential choices of the functions in (3)–(6) will be further discussed in the next section in light of the construction of a KF related to $\boldsymbol{f}_1(\cdot)$ and $\boldsymbol{f}_2(\cdot)$.

### 2.1. *The formulation of the closed-loop system*

In this paper, we want to construct a state feedback controller $\boldsymbol{u}(t) = K\boldsymbol{x}(t)$ with $K \in \mathbb{R}^{p\times n}$ to stabilize the open-loop system in (1). Then the corresponding closed-loop system is denoted as

$$\begin{aligned}
\dot{\boldsymbol{x}}(t) &= A_1\boldsymbol{x}(t) + \int_{-r_1}^{0} A_2\left(\widehat{\boldsymbol{f}}_1(\tau)\otimes I_n\right)\boldsymbol{x}(t+\tau)\mathsf{d}\tau + \int_{-r(t)}^{-r_1} A_3\left(\widehat{\boldsymbol{f}}_2(\tau)\otimes I_n\right)\boldsymbol{x}(t+\tau)\mathsf{d}\tau + B_1K\boldsymbol{x}(t)\\
&+ \int_{-r_1}^{0} B_2(I_{\kappa_1}\otimes K)\left(\widehat{\boldsymbol{f}}_1(\tau)\otimes I_n\right)\boldsymbol{x}(t+\tau)\mathsf{d}\tau + \int_{-r(t)}^{-r_1} B_3(I_{\kappa_2}\otimes K)\left(\widehat{\boldsymbol{f}}_2(\tau)\otimes I_n\right)\boldsymbol{x}(t+\tau)\mathsf{d}\tau + D_1\boldsymbol{w}(t)\\
\boldsymbol{z}(t) &= C_1\boldsymbol{x}(t) + \int_{-r_1}^{0} C_2\left(\widehat{\boldsymbol{f}}_1(\tau)\otimes I_n\right)\boldsymbol{x}(t+\tau)\mathsf{d}\tau + \int_{-r(t)}^{-r_1} C_3\left(\widehat{\boldsymbol{f}}_2(\tau)\otimes I_n\right)\boldsymbol{x}(t+\tau)\mathsf{d}\tau + B_4K\boldsymbol{x}(t)\\
&+ \int_{-r_1}^{0} B_5(I_{\kappa_1}\otimes K)\left(\widehat{\boldsymbol{f}}_1(\tau)\otimes I_n\right)\boldsymbol{x}(t+\tau)\mathsf{d}\tau + \int_{-r(t)}^{-r_1} B_6(I_{\kappa_2}\otimes K)\left(\widehat{\boldsymbol{f}}_2(\tau)\otimes I_n\right)\boldsymbol{x}(t+\tau)\mathsf{d}\tau + D_2\boldsymbol{w}(t),\\
&\forall\theta\in[-r_2,0],\quad \boldsymbol{x}(t_0+\theta)=\boldsymbol{\phi}(\tau),\quad r(\cdot)\in\mathbb{M}\left(\mathbb{R}\,⨾\,[r_1,r_2]\right)
\end{aligned}\tag{17}$$

by Lemma 2 and Proposition 1, where the decomposition of the distributed delays are constructed via

$$\left(\widehat{\boldsymbol{f}}_i(\tau)\otimes I_p\right)K = \left(\widehat{\boldsymbol{f}}_i(\tau)\otimes I_p\right)(1\otimes K) = \left(I_{\kappa_i}\widehat{\boldsymbol{f}}_i(\tau)\otimes KI_n\right) = (I_{\kappa_i}\otimes K)\left(\widehat{\boldsymbol{f}}_i(\tau)\otimes I_n\right),\quad i=1,2 \tag{18}$$

by (A.2). Note that (17) has different forms for the following three cases $r_2 > r_1 > 0$, and $r_1 = 0$; $r_2 > 0$, and $r_1 = r_2 > 0$.[2] This implies that each case of these three may require a distinct formulation for the corresponding synthesis conditions for (17). To avoid deriving three separated synthesis conditions, we rewrite (17) as

$$\begin{aligned}
&\dot{\boldsymbol{x}}(t) = \left(\mathbf{A}+\mathbf{B}_1\left[\left(I_{\widehat{3}+\kappa}\otimes K\right)\oplus \mathsf{O}_q\right]\right)\boldsymbol{\chi}(t),\quad \widetilde{\forall} t\geq t_0\\
&\boldsymbol{z}(t) = \left(\mathbf{C}+\mathbf{B}_2\left[\left(I_{\widehat{3}+\kappa}\otimes K\right)\oplus \mathsf{O}_q\right]\right)\boldsymbol{\chi}(t),\\
&\forall\theta\in[-r_2,0],\quad \boldsymbol{x}(t_0+\theta)=\boldsymbol{\phi}(\theta)
\end{aligned}\tag{19}$$

with $t_0$ and $\boldsymbol{\phi}(\cdot)$ in (1), where $\kappa = \kappa_1 + 2\kappa_2$ and

$$\mathbf{A} = \left[\widehat{\mathsf{O}}_{n,n}\quad A_1\quad A_2\left(\sqrt{\mathsf{G}_1}\otimes I_n\right)\quad A_3\left(\sqrt{\mathsf{G}_2}\otimes I_n\right)\quad \mathsf{O}_{n,\kappa_2 n}\quad D_1\right] \tag{20}$$

$$\mathbf{B}_1 = \left[\widehat{\mathsf{O}}_{n,p}\quad B_1\quad B_2\left(\sqrt{\mathsf{G}_1}\otimes I_p\right)\quad B_3\left(\sqrt{\mathsf{G}_2}\otimes I_p\right)\quad \mathsf{O}_{n,\kappa_2 p}\quad \mathsf{O}_{n,q}\right] \tag{21}$$

$$\mathbf{C} = \left[\widehat{\mathsf{O}}_{m,n}\quad C_1\quad C_2\left(\sqrt{\mathsf{G}_1}\otimes I_n\right)\quad C_3\left(\sqrt{\mathsf{G}_2}\otimes I_n\right)\quad \mathsf{O}_{m,\kappa_2 n}\quad D_2\right] \tag{22}$$

$$\mathbf{B}_2 = \left[\widehat{\mathsf{O}}_{m,p}\quad B_4\quad B_5\left(\sqrt{\mathsf{G}_1}\otimes I_p\right)\quad B_6\left(\sqrt{\mathsf{G}_2}\otimes I_p\right)\quad \mathsf{O}_{m,\kappa_2 p}\quad \mathsf{O}_{m,q}\right] \tag{23}$$

[2] Since (17) becomes a delay-free system with $r_1 = r_2 = 0$, hence such a case is not considered here.

$$
\boldsymbol{\chi}(t) = \begin{bmatrix} \widehat{\mathbb{1}}\boldsymbol{x}(t-r_1) \\ \mathbb{1}\boldsymbol{x}(t-r_2) \\ \boldsymbol{x}(t) \\ \int_{-r_1}^{0} \left( \sqrt{\mathsf{G}_1^{-1}}\widehat{\boldsymbol{f}}_1(\tau) \otimes I_n \right) \boldsymbol{x}(t+\tau)\mathsf{d}\tau \\ \int_{-r(t)}^{-r_1} \left( \sqrt{\mathsf{G}_2^{-1}}\widehat{\boldsymbol{f}}_2(\tau) \otimes I_n \right) \boldsymbol{x}(t+\tau)\mathsf{d}\tau \\ \int_{-r_2}^{-r(t)} \left( \sqrt{\mathsf{G}_2^{-1}}\widehat{\boldsymbol{f}}_2(\tau) \otimes I_n \right) \boldsymbol{x}(t+\tau)\mathsf{d}\tau \\ \boldsymbol{w}(t) \end{bmatrix}, \quad \begin{matrix} \widehat{\mathsf{O}}_{n,p} = \begin{cases} \mathsf{O}_{n,2p} & \text{for } \; r_2 > r_1 > 0 \\ \mathsf{O}_{n,p} & \text{for } \; r_1 = r_2 > 0 \\ \mathsf{O}_{n,p} & \text{for } \; r_1 = 0; r_2 > 0 \end{cases} \\ \widehat{3} = \begin{cases} 3 & \text{for } \; r_2 > r_1 > 0 \\ 2 & \text{for } \; r_1 = r_2 > 0 \\ 2 & \text{for } \; r_1 = 0; r_2 > 0 \end{cases} \end{matrix} \tag{24}
$$

$$
\mathbb{1} = \begin{cases} I_n & \text{for } \; r_2 > r_1 \geq 0 \\ [\,]_{0\times n} & \text{for } \; r_1 = r_2 > 0, \end{cases} \qquad \widehat{\mathbb{1}} = \begin{cases} I_n & \text{for } \; r_2 \geq r_1 > 0 \\ [\,]_{0\times n} & \text{for } \; r_1 = 0; r_2 > 0. \end{cases} \tag{25}
$$

Note that $\sqrt{X}$ stands for the unique square root of $X \succ 0$ and the terms in (20)–(23) are obtained by the following relations for $i \in \{1, 2\}$:

$$
\left( \widehat{\boldsymbol{f}}_i(\tau) \otimes I_n \right) = \sqrt{\mathsf{G}_i}\sqrt{\mathsf{G}_i^{-1}}\widehat{\boldsymbol{f}}_i(\tau) \otimes I_n = \left( \sqrt{\mathsf{G}_i} \otimes I_n \right) \left( \sqrt{\mathsf{G}_i^{-1}}\widehat{\boldsymbol{f}}_i(\tau) \otimes I_n \right), \tag{26}
$$

$$
(I_{\kappa_i} \otimes K) \left( \widehat{\boldsymbol{f}}_i(\tau) \otimes I_n \right) = \left( \sqrt{\mathsf{G}_i}\sqrt{\mathsf{G}_i^{-1}} \otimes K \right) \left( \widehat{\boldsymbol{f}}_i(\tau) \otimes I_n \right) = \left( \sqrt{\mathsf{G}_i} \otimes I_p \right) (I_{\kappa_i} \otimes K) \left( \sqrt{\mathsf{G}_i^{-1}}\widehat{\boldsymbol{f}}_i(\tau) \otimes I_n \right) \tag{27}
$$

which themselves can be obtained via (A.1) with the fact that $\mathsf{G}_1$ and $\mathsf{G}_2$ in (8) are invertible[3]. Now the expressions of the closed-loop system in (17) at $r_1 = r_2 > 0$ and $r_1 = 0$; $r_2 > 0$ can be obtained by (19) with $r_1 = r_2 > 0$, $d_2 = \delta_2 = 0$, and $r_1 = 0$; $r_2 > 0$, $d_1 = \delta_1 = 0$ in (20)–(24), respectively.

By introducing the terms $\widehat{\mathsf{O}}$, $\widehat{3}$, $\mathbb{1}$ and $\widehat{\mathbb{1}}$ in (24)–(25), the closed-loop system in (17) can be equivalently denoted by the form in (19) which can characterize all the cases of $r_2 \geq r_1 \geq 0$, $r_2 > 0$ without introducing redundant terms into the parameters in (20)–(24). This is critical in deriving well-posed synthesis conditions in this paper.

[3]Note that $\sqrt{X^{-1}} = \left(\sqrt{X}\right)^{-1}$ for any $X \succ 0$, based on the application of the eigendecomposition of $X \succ 0$

**Remark 3.** The existence and uniqueness of the solution of the closed-loop system (19) are guaranteed by Theorem 1.1 in Chapter 6 of Hale & Lunel (1993) which is developed for a general linear TDS. Specifically, consider $\int_{-r(t)}^{0} G(\tau)\boldsymbol{\phi}(\tau)\mathsf{d}\tau$ with $r(\cdot) \in \mathbb{M}\big(\mathbb{R} \,;\, [r_1, r_2]\big)$, $r_2 > 0$, $r_2 \geq r_1 \geq 0$ and $G(\cdot) \in \mathbb{L}^2([-r_2, 0] \,;\, \mathbb{R}^{m\times n})$ and $\boldsymbol{\phi}(\cdot) \in \mathbb{C}\left([-r_2, 0] \,;\, \mathbb{R}^n\right)$. By using the Cauchy Schwartz inequality with the fact that $G(\cdot) \in \mathbb{L}^2([-r_2, 0] \,;\, \mathbb{R}^{m\times n})$ and $\boldsymbol{\phi}(\cdot) \in \mathbb{C}\left([-r_2, 0] \,;\, \mathbb{R}^n\right) \subset \mathbb{L}^2\left([-r_2, 0] \,;\, \mathbb{R}^n\right)$, we have

$$
\begin{aligned}
&\left\| \int_{-r(t)}^{0} G(\tau)\boldsymbol{\phi}(\tau)\mathsf{d}\tau \right\|_2 = \left\| \int_{-r(t)}^{0} \mathop{\mathbf{Col}}_{i=1}^{m} \boldsymbol{g}_i^\top(\tau)\boldsymbol{\phi}(\tau)\mathsf{d}\tau \right\|_2 = \sqrt{\sum_{i=1}^{m} \left( \int_{-r(t)}^{0} \boldsymbol{g}_i^\top(\tau)\boldsymbol{\phi}(\tau)\mathsf{d}\tau \right)^2} \\
&\leq \sqrt{\sum_{i=1}^{m} \left( \int_{-r(t)}^{0} \|\boldsymbol{g}_i(\tau)\|_2^2 \,\mathsf{d}\tau \int_{-r(t)}^{0} \|\boldsymbol{\phi}(\tau)\|_2^2 \,\mathsf{d}\tau \right)} \leq \sqrt{\sum_{i=1}^{m} \left( \int_{-r_2}^{0} \|\boldsymbol{g}_i(\tau)\|_2^2 \,\mathsf{d}\tau \int_{-r_2}^{0} \|\boldsymbol{\phi}(\tau)\|_2^2 \,\mathsf{d}\tau \right)} \\
&\qquad\qquad \leq \sqrt{\sum_{i=1}^{m} \left( \alpha \int_{-r_2}^{0} \|\boldsymbol{\phi}(\cdot)\|_\infty^2 \,\mathsf{d}\tau \right)} = \sqrt{m r_2 \alpha \, \|\boldsymbol{\phi}(\cdot)\|_\infty^2} = \sqrt{m r_2 \alpha}\, \|\boldsymbol{\phi}(\cdot)\|_\infty
\end{aligned}
\tag{28}
$$

for some $\alpha > 0$, where $G(\tau) = \mathbf{Col}_{i=1}^{m}\, \boldsymbol{g}_i^\top(\tau)$. Now this shows that all the integrals in (19) satisfy the inequality below eq.(1.5) in Chapter 6 of Hale & Lunel (1993). This is because $\int_{-r(t)}^{0} G(\tau)\boldsymbol{\phi}(\tau)\mathsf{d}\tau = \int_{-r_2}^{0} \mathsf{u}(r(t)+\tau)G(\tau)\boldsymbol{\phi}(\tau)\mathsf{d}\tau$ that the function $\mathsf{u}(r(t)+\tau)G(\tau)$ is integrable in $\tau$ for all $t \in \mathbb{R}$ and measurable in $t \in \mathbb{R}$ for all $\tau \in [-r_2, 0]$.

## 3. Main results

Since the differential equation in (17) holds for almost all $t \geq t_0$ even in the case of $\boldsymbol{w}(t) \equiv \mathbf{0}_n$, thus the standard Lyapunov Krasovski stability theorem[4] cannot be applied to (17). A Lyapunov-Krasovski stability criterion is presented as follows which can analyze the stability of (17). See Lemma 4 in Appendix A for the general Lyapunov-Krasovski stability criterion which is derived for analyzing the stability of general functional differential equations subject to the Caratheodory conditions in section 2.6 of Hale & Lunel (1993).

**Corollary 1.** *Let $\boldsymbol{w}(t) \equiv \mathbf{0}_q$ in (19) and $r_2 \geq r_1 \geq 0$, $r_2 > 0$ be given, then the trivial solution $\boldsymbol{x}(t) \equiv \mathbf{0}_n$ of (19) is uniformly asymptotically stable in $\mathbb{C}([-r_2, 0] \,;\, \mathbb{R}^n)$ if there exist $\epsilon_1; \epsilon_2; \epsilon_3 > 0$ and a differentiable functional $\mathsf{v} : \mathbb{C}([-r_2, 0] \,;\, \mathbb{R}^n) \to \mathbb{R}$ with $\mathsf{v}(\mathsf{0}_n(\cdot)) = 0$ such that*

$$
\forall \boldsymbol{\phi}(\cdot) \in \mathbb{C}([-r_2, 0] \,;\, \mathbb{R}^n),\ \epsilon_1 \|\boldsymbol{\phi}(0)\|_2^2 \leq \mathsf{v}(\boldsymbol{\phi}(\cdot)) \leq \epsilon_2 \|\boldsymbol{\phi}(\cdot)\|_\infty^2 , \tag{29}
$$

$$
\widetilde{\forall} t \geq t_0 \in \mathbb{R},\ \frac{\mathsf{d}}{\mathsf{d}t}\mathsf{v}(\mathbf{x}_t(\cdot)) \leq -\epsilon_3 \|\boldsymbol{x}(t)\|_2^2 \tag{30}
$$

*where $\mathbf{x}_t(\cdot)$ in (30) is defined by the equality $\forall t \geq t_0$, $\forall \theta \in [-r_2, 0]$, $\mathbf{x}_t(\theta) = \boldsymbol{x}(t+\theta)$ in which $\boldsymbol{x}(\cdot) \in \mathbb{C}\left(\mathbb{R}_{\geq t_0 - r_2} \,;\, \mathbb{R}^n\right)$ satisfies $\dot{\boldsymbol{x}}(t) = \big(\mathbf{A} + \mathbf{B}_1 \left[\big(I_{\widehat{3+\kappa}} \otimes K\big) \oplus \mathsf{O}_q\right]\big)\boldsymbol{\chi}(t)$ in (19) with $\boldsymbol{w}(t) \equiv \mathbf{0}_q$.*

*Proof.* Since (19) with $\boldsymbol{w}(t) \equiv \mathbf{0}_q$ is a linear system and $r(\cdot) \in \mathbb{M}\left(\mathbb{R} \,;\, [r_1, r_2]\right)$, thus (19) with $\boldsymbol{w}(t) \equiv \mathbf{0}_q$ is a special case of the general time-varying system in (A.5). Then (29) and (30) can be obtained by letting $\alpha_1(s) = \epsilon_1 s^2$, $\alpha_2(s) = \epsilon_2 s^2$, $\alpha_3(s) = \epsilon_3 s^2$ with $\epsilon_1; \epsilon_2; \epsilon_3 > 0$. ■

[4] See Theorem 2.1 of Section 5.1 in Hale & Lunel (1993), and Theorem 1.3 in Gu *et al.* (2003)

**Definition 1.** Given $0 \neq r_2 \geq r_1 \geq 0$, the closed-loop system in (19) with a supply rate function $\mathsf{s}(\boldsymbol{z}(t), \boldsymbol{w}(t))$ is said to be dissipative if there exists a differentiable functional $\mathsf{v} : \mathbb{C}([-r_2, 0] ; \mathbb{R}^n) \to \mathbb{R}$ such that

$$\widetilde{\forall} t \geq t_0, \quad \dot{\mathsf{v}}(\mathbf{x}_t(\cdot)) - \mathsf{s}(\boldsymbol{z}(t), \boldsymbol{w}(t)) \leq 0 \tag{31}$$

where $t_0$, $\boldsymbol{z}(t)$ and $\boldsymbol{w}(t)$ are given in (19) together with $\forall t \geq t_0, \forall \theta \in [-r_2, 0]$, $\mathbf{x}_t(\theta) = \boldsymbol{x}(t+\theta)$ where $\boldsymbol{x}(t)$ is the solution of the system in (19).

Note that (31) implies the origin integral-based definition of dissipativity via the properties of Lebesgue integrals. To characterize dissipativity, a quadratic supply function

$$\mathsf{s}(\boldsymbol{z}(t), \boldsymbol{w}(t)) = \begin{bmatrix} \boldsymbol{z}(t) \\ \boldsymbol{w}(t) \end{bmatrix}^\top \begin{bmatrix} \widetilde{J}^\top J_1^{-1} \widetilde{J} & J_2 \\ * & J_3 \end{bmatrix} \begin{bmatrix} \boldsymbol{z}(t) \\ \boldsymbol{w}(t) \end{bmatrix}, \quad \mathbb{S}^m \ni \widetilde{J}^\top J_1^{-1} \widetilde{J} \preceq 0, \ \ \mathbb{S}^m \ni J_1^{-1} \prec 0, \ \ \widetilde{J} \in \mathbb{R}^{m \times m} \tag{32}$$

is applied in this paper where the structure of (32) is based the general quadratic constraints investigated in Scherer *et al.* (1997) together with the idea of factorizing the matrix $U_j$ in Scherer *et al.* (1997). Note that the supply rate function in (32) can characterize numerous performance criteria such as

- $\mathbb{L}^2$ gain performance: $J_1 = -\gamma I_m, \ \ \widetilde{J} = I_m, \ \ J_2 = \mathsf{O}_{m,q}, \ \ J_3 = \gamma I_q$ where $\gamma > 0$.
- Passivity: $J_1 \in \mathbb{S}^m_{\prec 0}, \ \ \widetilde{J} = \mathsf{O}_m, \ \ J_2 = I_m, \ \ J_3 = \mathsf{O}_m$ with $m = q$.

Two integral inequalities, which are presented in Lemma 5 and 6 in 5, are required to derive the main results in this paper. The second inequality in Lemma 6 is specifically proposed as an important contribution in this paper to ensure that the dimensions of the resulting synthesis conditions are finite.

The main results of this paper are summarized in two theorems and an algorithm in the rest of this section.

**Theorem 1.** *Let $r_2 > r_1 > 0$ and all the parameters in Proposition 1 be given, then the closed-loop system* (19) *with the supply rate function in* (32) *is dissipative and the trivial solution $\boldsymbol{x}(t) \equiv \mathbf{0}_n$ of* (19) *with $\boldsymbol{w}(t) \equiv \mathbf{0}_q$ is uniformly asymptotically stable in $\mathbb{C}([-r, 0] ; \mathbb{R}^n)$ if there exist $K \in \mathbb{R}^{p \times n}$ and $P_1 \in \mathbb{S}^n$, $P_2 \in \mathbb{R}^{n \times \varrho}$, $P_3 \in \mathbb{S}^\varrho$ with $\varrho = (d_1 + d_2)n$ and $Q_1; Q_2; R_1; R_2 \in \mathbb{S}^n$ and $Y \in \mathbb{R}^{n \times n}$ such that*

$$\begin{bmatrix} P_1 & P_2 \\ * & P_3 \end{bmatrix} + \left( \mathsf{O}_n \oplus \left[ I_{d_1} \otimes Q_1 \right] \oplus \left[ I_{d_2} \otimes Q_2 \right] \right) \succ 0, \tag{33}$$

$$Q_1 \succeq 0, \ Q_2 \succeq 0, \ R_1 \succeq 0, \ \begin{bmatrix} R_2 & Y \\ * & R_2 \end{bmatrix} \succeq 0, \tag{34}$$

$$\begin{bmatrix} \boldsymbol{\Psi} & \boldsymbol{\Sigma}^\top \widetilde{J}^\top \\ * & J_1 \end{bmatrix} = \mathbf{Sy} \left[ \mathbf{P}^\top \boldsymbol{\Pi} \right] + \boldsymbol{\Phi} \prec 0 \tag{35}$$

*where $\boldsymbol{\Sigma} = \mathbf{C} + \mathbf{B}_2 \left[ \left( I_{\widehat{3}+\kappa} \otimes K \right) \oplus \mathsf{O}_q \right]$ with $\mathbf{C}$ and $\mathbf{B}_2$ in* (22) *and* (23), *and*

$$\boldsymbol{\Psi} = \mathbf{Sy} \left( \begin{bmatrix} \widehat{\mathsf{O}}_{n,n}^\top & \widehat{\mathsf{O}}_{\varrho,n}^\top \\ I_n & \mathsf{O}_{n,\varrho} \\ \mathsf{O}_{\kappa n,n} & \widehat{I}^\top \\ \mathsf{O}_{q,n} & \mathsf{O}_{q,\varrho} \end{bmatrix} \begin{bmatrix} P_1 & P_2 \\ * & P_3 \end{bmatrix} \begin{bmatrix} \mathbf{A} + \mathbf{B}_1 \left[ \left( I_{\widehat{3}+\kappa} \otimes K \right) \oplus \mathsf{O}_q \right] \\ \begin{bmatrix} \widehat{\mathbf{F}} \otimes I_n & \mathsf{O}_{\varrho,q} \end{bmatrix} \end{bmatrix} - \begin{bmatrix} \widehat{\mathsf{O}}_{m,n}^\top \\ \mathsf{O}_{(n+\kappa n),m} \\ J_2^\top \end{bmatrix} \boldsymbol{\Sigma} \right) - \Xi \tag{36}$$

$$\widehat{I} = \left( \sqrt{\mathsf{F}_1^{-1}} \oplus \sqrt{\mathsf{F}_2^{-1}} \right) \begin{bmatrix} \mathsf{O}_{d_1,\delta_1} & I_{d_1} & \mathsf{O}_{d_1,\delta_2} & \mathsf{O}_{d_1,d_2} & \mathsf{O}_{d_1,\delta_2} & \mathsf{O}_{d_1,d_2} \\ \mathsf{O}_{d_2,\delta_1} & \mathsf{O}_{d_2,d_1} & \mathsf{O}_{d_2,\delta_2} & I_{d_2} & \mathsf{O}_{d_2,\delta_2} & I_{d_2} \end{bmatrix} \left( \sqrt{\mathsf{G}_1} \oplus \sqrt{\mathsf{G}_2} \oplus \sqrt{\mathsf{G}_2} \right) \otimes I_n \tag{37}$$

$$
\begin{aligned}
\Xi = \Big[ & [Q_1 - Q_2 - r_3 R_2] \oplus [\mathbb{1} Q_2] \oplus \left[\widehat{\mathbb{1}}(-Q_1 - r_1 R_1)\right] \oplus (I_{\kappa_1} \otimes R_1) \\
& \oplus \left( \begin{bmatrix} \mathsf{K}_{(\kappa_2,n)} & \mathsf{O}_{\kappa_2 n} \\ * & \mathsf{K}_{(\kappa_2,n)} \end{bmatrix} \left( \begin{bmatrix} R_2 & Y \\ * & R_2 \end{bmatrix} \otimes I_{\kappa_2} \right) \begin{bmatrix} \mathsf{K}_{(n,\kappa_2)} & \mathsf{O}_{\kappa_2 n} \\ * & \mathsf{K}_{(n,\kappa_2)} \end{bmatrix} \right) \oplus J_3 \Big]
\end{aligned} \tag{38}
$$

$$
\widehat{\mathbf{F}} = \begin{bmatrix} -\sqrt{\mathsf{F}_1^{-1}} \boldsymbol{f}_1(-r_1) & \mathbf{0}_{d_1} & \sqrt{\mathsf{F}_1^{-1}} \boldsymbol{f}_1(0) & -\sqrt{\mathsf{F}_1^{-1}} M_1 \sqrt{\mathsf{G}_1} & \mathsf{O}_{d_1,\kappa_2} & \mathsf{O}_{d_1,\kappa_2} \\ \sqrt{\mathsf{F}_2^{-1}} \boldsymbol{f}_2(-r_1) & -\sqrt{\mathsf{F}_2^{-1}} \boldsymbol{f}_2(-r_2) & \mathbf{0}_{d_2} & \mathsf{O}_{d_2,\kappa_1} & -\sqrt{\mathsf{F}_2^{-1}} M_2 \sqrt{\mathsf{G}_2} & -\sqrt{\mathsf{F}_2^{-1}} M_2 \sqrt{\mathsf{G}_2} \end{bmatrix} \tag{39}
$$

*with* $\mathbf{A}$, $\mathbf{B}_1$ *in* (20)–(21) *and* $\mathbb{1}$, $\widehat{\mathbb{1}}$ *in* (25) *and* $\mathsf{G}_1$, $\mathsf{G}_2$ *in* (8)–(9). *Moreover,* $\mathsf{F}_1 = \int_{-r_1}^{0} \boldsymbol{f}_1(\tau) \boldsymbol{f}_1^\top(\tau) \mathsf{d}\tau$ *and* $\mathsf{F}_2 = \int_{-r_2}^{-r_1} \boldsymbol{f}_2(\tau) \boldsymbol{f}_2^\top(\tau) \mathsf{d}\tau$ *and the rest of the parameters in* (35) *is defined as*

$$
\mathbf{P} = \left[ \widehat{\mathsf{O}}_{n,n} \quad P_1 \quad P_2 \widehat{I} \quad \mathsf{O}_{n,q} \quad \mathsf{O}_{n,m} \right], \quad \mathbf{\Pi} = \left[ \mathbf{A} + \mathbf{B}_1 \left[ \left( I_{\widehat{3}+\kappa} \otimes K \right) \oplus \mathsf{O}_q \right] \quad \mathsf{O}_{n,m} \right] \tag{40}
$$

*and*

$$
\mathbf{\Phi} = \mathbf{Sy} \left( \begin{bmatrix} \widehat{\mathsf{O}}_{\varrho,n}^\top \\ P_2 \\ \widehat{I}^\top P_3 \\ \mathsf{O}_{(q+m),\varrho} \end{bmatrix} \left[ \widehat{\mathbf{F}} \otimes I_n \quad \mathsf{O}_{\varrho,(q+m)} \right] + \begin{bmatrix} \widehat{\mathsf{O}}_{m,n}^\top \\ \mathsf{O}_{(n+\kappa n),m} \\ -J_2^\top \\ \widetilde{J} \end{bmatrix} \left[ \mathbf{\Sigma} \quad \mathsf{O}_m \right] \right) - \Xi \oplus (-J_1) \, . \tag{41}
$$

*Furthermore, with* $r_1 = r_2$, $d_2 = \delta_2 = 0$ *and* $Q_2 = R_2 = Y = \mathsf{O}_n$, *then the inequalities in* (33)–(35) *are a dissipative synthesis condition for the closed-loop system in* (19) *with* $r_1 = r_2 > 0$. *Finally, with* $r_2 > 0$; $r_1 = 0$, $d_1 = \delta_1 = 0$ *and* $Q_1 = R_1 = \mathsf{O}_n$, *then the inequalities in* (33)–(35) *are a dissipative synthesis condition for the closed-loop system in* (19) *with* $r_2 > 0$; $r_1 = 0$.

*Proof.* See Appendix C. ∎

**Remark 4.** Without using $\mathbb{1}$, $\widehat{\mathbb{1}}$ and $\widehat{\mathsf{O}}$, the synthesis condition derived for the case of $r_2 > r_1 > 0$ may not be directly applied to the cases of $r_1 = r_2$ or $r_1 = 0; r_2 > 0$. This is due to the changes of the mathematical structures of the closed-loop system in (19) and the functional (C.1) corresponding to $r_1 = r_2$ or $r_1 = 0; r_2 > 0$.

**Remark 5.** Note that $\boldsymbol{f}_1(\cdot)$ and $\boldsymbol{f}_2(\cdot)$ in (C.2) can be any differentiable function since the decompositions in (3)–(6) are always constructible via some proper choices of $\varphi_1(\cdot)$ and $\varphi_2(\cdot)$. This provides great flexibility to the structure of the Liapunov-Krasovski functional in (C.1). On the other hand, the functions inside of $\boldsymbol{f}_1(\cdot)$ and $\boldsymbol{f}_2(\cdot)$ can be chosen in view of the functions inside of the distributed delays in (1).

### 3.1. *A comment on* (B.5)

The significance of the proposed inequality in (B.5) can be understood considering the procedures in the proof of Theorem 1. Indeed, assume that (B.2) is directly applied to the integrals $\int_{-r(t)}^{-r_1} \boldsymbol{x}^\top(t+\tau) Q_2 \boldsymbol{x}(t+\tau) \mathsf{d}\tau$ and $\int_{-r_2}^{-r(t)} \boldsymbol{x}^\top(t+\tau) Q_2 \boldsymbol{x}(t+\tau) \mathsf{d}\tau$ without using (B.5) at the step in (C.9), which gives the inequalities

$$
\begin{aligned}
\int_{-r(t)}^{-r_1} \boldsymbol{x}^\top(t+\tau) Q_2 \boldsymbol{x}(t+\tau) \mathsf{d}\tau &\geq [*] \left( \widehat{\mathsf{F}}_1^{-1}(r(t)) \otimes Q_2 \right) \left[ \int_{-r(t)}^{-r_1} \left( \widehat{\boldsymbol{f}}_2(\tau) \otimes I_n \right) \boldsymbol{x}(t+\tau) \mathsf{d}\tau \right] \\
\int_{-r_2}^{-r(t)} \boldsymbol{x}^\top(t+\tau) Q_2 \boldsymbol{x}(t+\tau) \mathsf{d}\tau &\geq [*] \left( \widehat{\mathsf{F}}_2^{-1}(r(t)) \otimes Q_2 \right) \left[ \int_{-r_2}^{-r(t)} \left( \widehat{\boldsymbol{f}}_2(\tau) \otimes I_n \right) \boldsymbol{x}(t+\tau) \mathsf{d}\tau \right]
\end{aligned} \tag{42}
$$

where $\widehat{\mathsf{F}}_1(r(t)) = \int_{-r(t)}^{-r_1} \widehat{\boldsymbol{f}}_2(\tau)\widehat{\boldsymbol{f}}_2^\top(\tau)\mathsf{d}\tau$ and $\widehat{\mathsf{F}}_2(r(t)) = \int_{-r_2}^{-r(t)} \widehat{\boldsymbol{f}}_2(\tau)\widehat{\boldsymbol{f}}_2^\top(\tau)\mathsf{d}\tau$. Now combine the inequalities in (42), we have

$$\int_{-r_2}^{-r_1} \boldsymbol{x}^\top(t+\tau)Q_2\boldsymbol{x}(t+\tau)\mathsf{d}\tau \geq \begin{bmatrix} \int_{-r(t)}^{-r_1} \left(\widehat{\boldsymbol{f}}_2(\tau)\otimes I_n\right)\boldsymbol{x}(t+\tau)\mathsf{d}\tau \\ \int_{-r_2}^{-r(t)} \left(\widehat{\boldsymbol{f}}_2(\tau)\otimes I_n\right)\boldsymbol{x}(t+\tau)\mathsf{d}\tau \end{bmatrix}^\top \times$$

$$\begin{bmatrix} \widehat{\mathsf{F}}_1^{-1}(r(t))\otimes Q_2 & \mathsf{O}_{d_1n,d_2n} \\ \mathsf{O}_{d_2n,d_1n} & \widehat{\mathsf{F}}_2^{-1}(r(t))\otimes Q_2 \end{bmatrix} \begin{bmatrix} \int_{-r(t)}^{-r_1} \left(\widehat{\boldsymbol{f}}_2(\tau)\otimes I_n\right)\boldsymbol{x}(t+\tau)\mathsf{d}\tau \\ \int_{-r_2}^{-r(t)} \left(\widehat{\boldsymbol{f}}_2(\tau)\otimes I_n\right)\boldsymbol{x}(t+\tau)\mathsf{d}\tau \end{bmatrix} \tag{43}$$

which also furnishes a lower bound for $\int_{-r_2}^{-r_1} \boldsymbol{x}^\top(t+\tau)Q_2\boldsymbol{x}(t+\tau)\mathsf{d}\tau$. Conventionally, the reciprocally convex combination lemma Park *et al.* (2011) or its derivatives Seuret & Gouaisbaut (2017); Zhang *et al.* (2017a,b) can be applied to a matrix in the form of $\begin{bmatrix} \frac{1}{1-\alpha}X & \mathsf{O}_n \\ \mathsf{O}_n & \frac{1}{\alpha}X \end{bmatrix}$ to construct a tractable lower bound with finite dimensions. However, the structure of $\begin{bmatrix} \frac{1}{1-\alpha}X & \mathsf{O}_n \\ \mathsf{O}_n & \frac{1}{\alpha}X \end{bmatrix}$ may not be always guaranteed by the matrix

$$\begin{bmatrix} \mathsf{F}_1^{-1}(r(t))\otimes Q_2 & \mathsf{O}_{d_1n,d_2n} \\ \mathsf{O}_{d_2n,d_1n} & \mathsf{F}_2^{-1}(r(t))\otimes Q_2 \end{bmatrix} \tag{44}$$

in (43), since $\mathsf{F}_1^{-1}(r(t))$ and $\mathsf{F}_2^{-1}(r(t))$ are nonlinear with respect to $r(t)$ in general.[5] On the other hand, if (43) is applied directly to replace the step at (C.8) without the use of any kind of reciprocally convex combination lemmas, then the matrix in (44) will appear in the corresponding (35), where (35) becomes infinite-dimensional and also generally nonlinear with respect to $r(t)$. In contrast, the symmetric matrix in the lower bound in (C.9) is of finite-dimensional, which is constructed via the application of (B.5). This shows the contribution of the integral inequality in (B.5) by which a dissipative synthesis condition with finite dimensions can be derived via the Krasovski functional method.

### 3.2. *A convex dissipative synthesis condition*

$\mathbf{Sy}\left[\mathbf{P}^\top\mathbf{\Pi}\right] + \mathbf{\Phi} \prec 0$ in (35) is bilinear with respect to the variables in $\mathbf{P}$ and $\mathbf{\Pi}$ if a synthesis problem is concerned, which cannot be solved directly via standard SDP solvers. To tackle this problem, a convex dissipative synthesis condition is constructed in the following theorem via the application of Projection Lemma Gahinet & Apkarian (1994) to (35).

**Lemma 1** (Projection Lemma)**.** *Gahinet & Apkarian (1994) Given $n; p; q \in \mathbb{N}$, $\Pi \in \mathbb{S}^n, P \in \mathbb{R}^{q\times n}, Q \in \mathbb{R}^{p\times n}$, there exists $\Theta \in \mathbb{R}^{p\times q}$ such that the following two propositions are equivalent :*

$$\Pi + P^\top\Theta^\top Q + Q^\top\Theta P \prec 0, \tag{45}$$

$$P_\perp^\top \Pi P_\perp \prec 0 \ \ and \ \ Q_\perp^\top \Pi Q_\perp \prec 0, \tag{46}$$

*where the columns of $P_\perp$ and $Q_\perp$ contain bases of null space of matrix $P$ and $Q$, respectively, which means that $PP_\perp = \mathsf{O}$ and $QQ_\perp = \mathsf{O}$.*

*Proof.* Refer to the proof of Lemma 3.1 of Gahinet & Apkarian (1994) and Lemma C.12.1 of Briat (2014). ■

[5]If $\widehat{\boldsymbol{f}}_2(\tau)$ only contains Legendre polynomials with appropriate structures, then the reciprocally convex combination lemma or its derivatives can be applied to (44). Nevertheless, this is a very special case of $\widehat{\boldsymbol{f}}_2(\cdot) \in \mathbb{L}^2\left([-r_2, 0] ; \mathbb{R}^{d_2+\delta_2}\right)$ considered in this paper.

**Theorem 2.** *Given* $\{\alpha_i\}_{i=1}^{\widehat{3}+\kappa} \subset \mathbb{R}$ *and* $r_2 > r_1 > 0$ *and the functions and parameters in Proposition 1, then the closed-loop system in* (19) *with the supply rate function in* (32) *is dissipative and the trivial solution* $\boldsymbol{x}(t) \equiv \mathbf{0}_n$ *of* (19) *with* $\boldsymbol{w}(t) \equiv \mathbf{0}_q$ *is uniformly asymptotically stable in* $\mathbb{C}([-r, 0] ; \mathbb{R}^n)$ *if there exists* $\acute{P}_1 \in \mathbb{S}^n$, $\acute{P}_2 \in \mathbb{R}^{n\times\varrho}$, $\acute{P}_3 \in \mathbb{S}^{\varrho}$ *and* $\acute{Q}_1; \acute{Q}_2; \acute{R}_1; \acute{R}_2; X \in \mathbb{S}^n$ *and* $\acute{Y} \in \mathbb{R}^{n\times n}$ *and* $V \in \mathbb{R}^{p\times n}$ *such that*

$$\begin{bmatrix} \acute{P}_1 & \acute{P}_2 \\ * & \acute{P}_3 \end{bmatrix} + \left(\mathsf{O}_n \oplus \left[I_{d_1} \otimes \acute{Q}_1\right] \oplus \left[I_{d_2} \otimes \acute{Q}_2\right]\right) \succ 0, \tag{47}$$

$$\acute{Q}_1 \succeq 0,\ \acute{Q}_2 \succeq 0,\ \acute{R}_1 \succeq 0,\ \begin{bmatrix} \acute{R}_2 & \acute{Y} \\ * & \acute{R}_2 \end{bmatrix} \succeq 0, \tag{48}$$

$$\mathbf{Sy}\left(\begin{bmatrix} I_n \\ \mathbf{Col}_{i=1}^{\widehat{3}+\kappa}\, \alpha_i I_n \\ \mathsf{O}_{(q+m),n} \end{bmatrix} \begin{bmatrix} -X & \acute{\mathbf{\Pi}} \end{bmatrix}\right) + \begin{bmatrix} \mathsf{O}_n & \acute{\mathbf{P}} \\ * & \acute{\mathbf{\Phi}} \end{bmatrix} \prec 0 \tag{49}$$

*where* $\acute{\mathbf{\Pi}} = \left[\mathbf{A}\left[\left(I_{\widehat{3}+\kappa} \otimes X\right) \oplus I_q\right] + \mathbf{B}_1\left[\left(I_{\widehat{3}+\kappa} \otimes V\right) \oplus \mathsf{O}_q\right] \quad \mathsf{O}_{n,m}\right]$ *and* $\acute{\mathbf{P}} = \left[\widehat{\mathsf{O}}_{n,n} \quad \acute{P}_1 \quad \acute{P}_2\widehat{I} \quad \mathsf{O}_{n,(q+m)}\right]$ *with* $\widehat{I}$ *in* (37) *and*

$$\begin{aligned} \acute{\mathbf{\Phi}} = \mathbf{Sy}&\left(\begin{bmatrix} \widehat{\mathsf{O}}_{\varrho,n}^{\top} \\ \acute{P}_2 \\ \widehat{I}^{\top}\acute{P}_3 \\ \mathsf{O}_{(q+m),\varrho} \end{bmatrix} \begin{bmatrix} \widehat{\mathbf{F}} \otimes I_n & \mathsf{O}_{\varrho,(q+m)} \end{bmatrix} + \begin{bmatrix} \widehat{\mathsf{O}}_{m,n}^{\top} \\ \mathsf{O}_{(n+\kappa n),m} \\ -J_2^{\top} \\ \widetilde{J} \end{bmatrix} \begin{bmatrix} \acute{\mathbf{\Sigma}} & \mathsf{O}_m \end{bmatrix}\right) \\ &- \Bigg(\left[\acute{Q}_1 - \acute{Q}_2 - r_3\acute{R}_2\right] \oplus \mathbb{1}\acute{Q}_2 \oplus \left[\widehat{\mathbb{1}}(-\acute{Q}_1 - r_1\acute{R}_1)\right] \oplus \left[I_{\kappa_1} \otimes \acute{R}_1\right] \\ &\oplus \left([*]\left(\begin{bmatrix} \acute{R}_2 & \acute{Y} \\ * & \acute{R}_2 \end{bmatrix} \otimes I_{\kappa_2}\right)\begin{bmatrix} \mathsf{K}_{(n,\kappa_2)} & \mathsf{O}_{\kappa_2 n} \\ * & \mathsf{K}_{(n,\kappa_2)} \end{bmatrix}\right) \oplus J_3 \oplus (-J_1)\Bigg) \end{aligned} \tag{50}$$

*with* $\acute{\mathbf{\Sigma}} = \mathbf{C}\left[\left(I_{\widehat{3}+\kappa} \otimes X\right) \oplus I_q\right] + \mathbf{B}_2\left[\left(I_{\widehat{3}+\kappa} \otimes V\right) \oplus \mathsf{O}_q\right]$ *and* $\mathbf{A}, \mathbf{B}_1, \mathbf{B}_2, \mathbf{C}$ *are given in* (20)–(23). *The controller gain is calculated via* $K = VX^{-1}$. *Furthermore, with* $r_1 = r_2$, $d_2 = \delta_2 = 0$ *and* $\acute{Q}_2 = \acute{R}_2 = \acute{Y} = \mathsf{O}_n$, *then the inequalities in* (47)–(49) *are a dissipative synthesis condition for the closed-loop system with* $r_1 = r_2 > 0$. *Finally, with* $r_2 > 0$; $r_1 = 0$, $d_1 = \delta_1 = 0$ *and* $\acute{Q}_1 = \acute{R}_1 = \mathsf{O}_n$, *then the inequalities in* (47)–(49) *are a dissipative synthesis condition for the closed-loop system with* $r_2 > 0$; $r_1 = 0$.

*Proof.* Consider the case of $r_2 > r_1 > 0$. First of all, note that the inequality $\mathbf{Sy}\left(\mathbf{P}^{\top}\mathbf{\Pi}\right) + \mathbf{\Phi} \prec 0$ in (35) can be reformulated into

$$\mathbf{Sy}\left(\mathbf{P}^{\top}\mathbf{\Pi}\right) + \mathbf{\Phi} = \begin{bmatrix} \mathbf{\Pi} \\ I_{3n+\kappa n+q+m} \end{bmatrix}^{\top} \begin{bmatrix} \mathsf{O}_n & \mathbf{P} \\ * & \mathbf{\Phi} \end{bmatrix} \begin{bmatrix} \mathbf{\Pi} \\ I_{3n+\kappa n+q+m} \end{bmatrix} \prec 0. \tag{51}$$

where the structure of (51) is similar to one of the inequalities in (46) as part of the statements of Lemma 1. Given the fact that there are two matrix inequalities in (46), thus a new matrix inequality must be constructed accordingly to use Lemma 1 in order to decouple the product between $\mathbf{P}$ and $\mathbf{\Pi}$ in (51). Now consider

$$\Upsilon^{\top} \begin{bmatrix} \mathsf{O}_n & \mathbf{P} \\ * & \mathbf{\Phi} \end{bmatrix} \Upsilon \prec 0 \tag{52}$$

with $\Upsilon^{\top} := \left[\mathsf{O}_{(q+m),(4n+\kappa n)} \quad I_{q+m}\right]$. The inequality in (52) can be further simplified as

$$\Upsilon^{\top} \begin{bmatrix} \mathsf{O}_n & \mathbf{P} \\ * & \mathbf{\Phi} \end{bmatrix} \Upsilon = \begin{bmatrix} -J_3 - \mathbf{Sy}(D_2^{\top}J_2) & D_2^{\top}\widetilde{J} \\ * & J_1 \end{bmatrix} \prec 0. \tag{53}$$

where the left-hand side of the inequality in (53) is the $2\times 2$ block matrix at the right bottom of $\mathbf{Sy}\left(\mathbf{P}^\top\mathbf{\Pi}\right)+\mathbf{\Phi}$ or $\mathbf{\Phi}$. As a result, it is clear that (53) is automatically satisfied if (51) or (35) holds. Hence (53) and (35) hold if and only if (35) holds. On the other hand, the following identities

$$\begin{gathered}\begin{bmatrix}-I_n & \mathbf{\Pi}\end{bmatrix}\begin{bmatrix}\mathbf{\Pi}\\ I_{3n+\kappa n+q+m}\end{bmatrix}=\mathsf{O}_{n,(3n+\kappa n+q+m)},\quad \begin{bmatrix}-I_n & \mathbf{\Pi}\end{bmatrix}_\perp=\begin{bmatrix}\mathbf{\Pi}\\ I_{3n+\kappa n+q+m}\end{bmatrix}\\ \begin{bmatrix}I_{4n+\kappa n} & \mathsf{O}_{(4n+\kappa n),(q+m)}\end{bmatrix}\begin{bmatrix}\mathsf{O}_{(4n+\kappa n),(q+m)}\\ I_{q+m}\end{bmatrix}=\begin{bmatrix}I_{4n+\kappa n} & \mathsf{O}_{(4n+\kappa n),(q+m)}\end{bmatrix}\Upsilon=\mathsf{O}_{(4n+\kappa n),(q+m)}\\ \begin{bmatrix}I_{4n+\kappa n} & \mathsf{O}_{(4n+\kappa n),(q+m)}\end{bmatrix}_\perp=\begin{bmatrix}\mathsf{O}_{(4n+\kappa n),(q+m)}\\ I_{q+m}\end{bmatrix}=\Upsilon\end{gathered}\tag{54}$$

where $\operatorname{rank}\left(\begin{bmatrix}-I_n & \mathbf{\Pi}\end{bmatrix}\right)=n$ and $\operatorname{rank}\left(\begin{bmatrix}I_{4n+\kappa n} & \mathsf{O}_{(4n+\kappa n),(q+m)}\end{bmatrix}\right)=4n+\kappa n$, imply that Lemma 1 can be used with the terms in (54) given the rank nullity theorem.

Applying Lemma 1 to (51) and (53) with (54) yields the conclusion that (51) holds if and only if

$$\exists\mathbf{W}\in\mathbb{R}^o:\mathbf{Sy}\left(\begin{bmatrix}I_{4n+\kappa n}\\ \mathsf{O}_{(q+m),(4n+\kappa n)}\end{bmatrix}\mathbf{W}\begin{bmatrix}-I_n & \mathbf{\Pi}\end{bmatrix}\right)+\begin{bmatrix}\mathsf{O}_n & \mathbf{P}\\ * & \mathbf{\Phi}\end{bmatrix}\prec 0.\tag{55}$$

Now the inequality in (55) is still bilinear due to the product between $\mathbf{W}$ and $\mathbf{\Pi}$. To convexify (55), consider

$$\mathbf{W}:=\mathbf{Col}\left[W,\ \mathbf{Col}_{i=1}^{\widehat{3}+\kappa}\alpha_i W\right]\tag{56}$$

with $W\in\mathbb{S}^n$ and $\{\alpha_i\}_{i=1}^{\widehat{3}+\kappa}\subset\mathbb{R}$. With (56), the inequality in (55) becomes

$$\mathbf{\Theta}=\mathbf{Sy}\left(\begin{bmatrix}W\\ \mathbf{Col}_{i=1}^{3+\kappa}\alpha_i W\\ \mathsf{O}_{(q+m),n}\end{bmatrix}\begin{bmatrix}-I_n & \mathbf{\Pi}\end{bmatrix}\right)+\begin{bmatrix}\mathsf{O}_n & \mathbf{P}\\ * & \mathbf{\Phi}\end{bmatrix}\prec 0\tag{57}$$

which infers (51). Note that using the structured in (56) infers that (57) is no longer an equivalent but only a sufficient condition implying (51) which is equivalent to (35). It is also important to stress that an invertible $W$ is automatically inferred by (57) since the expression $-2W$ is the only element at the first top left diagonal block of $\mathbf{\Theta}$.

Let $X^\top=W^{-1}$, we now apply congruence transformations (Caverly & Forbes, 2019, page 12) to the matrix inequalities in (33),(34) and (57) with the fact that an invertible $W$ is inferred by (57). Then one can conclude that

$$\begin{gathered}X^\top Q_1X\succ 0,\ \ X^\top Q_2X\succ 0,\ \ X^\top R_1X\succ 0,\ \begin{bmatrix}X^\top & \mathsf{O}_n\\ * & X^\top\end{bmatrix}\begin{bmatrix}R_2 & Y\\ * & R_2\end{bmatrix}\begin{bmatrix}X & \mathsf{O}_n\\ * & X\end{bmatrix}\succ 0,\\ \left[\left(I_{4+\kappa}\otimes X^\top\right)\oplus I_{q+m}\right]\mathbf{\Theta}\left[\left(I_{4+\kappa}\otimes X\right)\oplus I_{q+m}\right]\prec 0,\\ [*]\left(\begin{bmatrix}P_1 & P_2\\ * & P_3\end{bmatrix}+\left(\mathsf{O}_n\oplus\left[I_{d_1}\otimes Q_1\right]\oplus\left[I_{d_2}\otimes Q_2\right]\right)\right)\left(I_{1+d_1+d_2}\otimes X\right)\succ 0\end{gathered}\tag{58}$$

hold if and only if (33),(34) and (57) hold. Moreover, considering (A.1) and the definitions $\acute{Y}:=X^\top YX$ and

$$\begin{bmatrix}\acute{P}_1 & \acute{P}_2\\ * & \acute{P}_3\end{bmatrix}:=[*]\begin{bmatrix}P_1 & P_2\\ * & P_3\end{bmatrix}\left(I_{1+d_1+d_2}\otimes X\right),\ \ \begin{bmatrix}\acute{Q}_1 & \acute{Q}_2 & \acute{R}_1 & \acute{R}_2\end{bmatrix}:=X^\top\begin{bmatrix}Q_1X & Q_2X & R_1X & R_2X\end{bmatrix},\tag{59}$$

then the inequalities in (58) can be rewritten into (47) and (48) and

$$[*]\,\mathbf{\Theta}\left[\left(I_{4+\kappa}\otimes X\right)\oplus I_{q+m}\right]=\acute{\mathbf{\Theta}}=\mathbf{Sy}\left(\begin{bmatrix}I_n\\ \mathbf{Col}_{i=1}^{3+\kappa}\alpha_i I_n\\ \mathsf{O}_{(q+m),n}\end{bmatrix}\begin{bmatrix}-X & \acute{\mathbf{\Pi}}\end{bmatrix}\right)+\begin{bmatrix}\mathsf{O}_n & \acute{\mathbf{P}}\\ * & \acute{\mathbf{\Phi}}\end{bmatrix}\prec 0\tag{60}$$

where $\acute{\mathbf{P}} = X\mathbf{P}\left[(I_{3+\kappa} \otimes X) \oplus I_{q+m}\right] = \begin{bmatrix}\widehat{\mathrm{O}}_{n,n} & \acute{P}_1 & \acute{P}_2\widehat{I} & \mathsf{O}_{n,q} & \mathsf{O}_{n,m}\end{bmatrix}$ and

$$
\begin{aligned}
\acute{\mathbf{\Pi}} = \mathbf{\Pi}\left[(I_{3+\kappa} \otimes X) \oplus I_{q+m}\right] &= \begin{bmatrix}\mathbf{A}\left[(I_{3+\kappa} \otimes X) \oplus I_q\right] + \mathbf{B}_1\left[(I_{3+\kappa} \otimes KX) \oplus \mathsf{O}_q\right] & \mathsf{O}_{n,m}\end{bmatrix} \\
&= \begin{bmatrix}\mathbf{A}\left[(I_{3+\kappa} \otimes X) \oplus I_q\right] + \mathbf{B}_1\left[(I_{3+\kappa} \otimes V) \oplus \mathsf{O}_q\right] & \mathsf{O}_{n,m}\end{bmatrix} \qquad (61)
\end{aligned}
$$

with $V = KX$ and $\acute{\mathbf{\Phi}}$ in (50). Note that (60)–(61) is equivalent to the statements in Theorem 2 given the definition of $\widehat{3}$ and $\widehat{\mathrm{O}}$ in (24). Note that also the form of $\acute{\mathbf{\Phi}}$ in (50) is derived via the relations $\widehat{I}\left(I_\kappa \otimes X\right) = \left(I_{d_1+d_2} \otimes X\right)\widehat{I}$ and

$$
\begin{aligned}
&\begin{bmatrix}\widehat{\mathbf{F}} \otimes I_n & \mathsf{O}_{\varrho,(q+m)}\end{bmatrix}\left[(I_{3+\kappa} \otimes X) \oplus I_{q+m}\right] = \begin{bmatrix}I_{d_1+d_2}\widehat{\mathbf{F}} \otimes XI_n & \mathsf{O}_{\varrho,(q+m)}\end{bmatrix} \\
&\quad = \begin{bmatrix}\left(I_{d_1+d_2} \otimes X\right)\left(\widehat{\mathbf{F}} \otimes I_n\right) & \mathsf{O}_{\varrho,(q+m)}\end{bmatrix} = \left(I_{d_1+d_2} \otimes X\right)\begin{bmatrix}\widehat{\mathbf{F}} \otimes I_n & \mathsf{O}_{\varrho,(q+m)}\end{bmatrix},
\end{aligned} \qquad (62)
$$

$$
\begin{aligned}
\begin{bmatrix}\mathsf{K}_{(n,\kappa_2)} & \mathsf{O}_{\kappa_2 n} \\ * & \mathsf{K}_{(n,\kappa_2)}\end{bmatrix}\begin{bmatrix}I_{\kappa_2} \otimes X & \mathsf{O}_{\kappa_2 n} \\ * & I_{\kappa_2} \otimes X\end{bmatrix} &= \begin{bmatrix}X \otimes I_{\kappa_2} & \mathsf{O}_{\kappa_2 n} \\ * & X \otimes I_{\kappa_2}\end{bmatrix}\begin{bmatrix}\mathsf{K}_{(n,\kappa_2)} & \mathsf{O}_{\kappa_2 n} \\ * & \mathsf{K}_{(n,\kappa_2)}\end{bmatrix} \\
&= \left(\begin{bmatrix}X & \mathsf{O}_n \\ * & X\end{bmatrix} \otimes I_{\kappa_2}\right)\begin{bmatrix}\mathsf{K}_{(n,\kappa_2)} & \mathsf{O}_{\kappa_2 n} \\ * & \mathsf{K}_{(n,\kappa_2)}\end{bmatrix}
\end{aligned} \qquad (63)
$$

which are derived from the properties of matrices with (A.1),(A.2) and (A.3). Furthermore, since $-2X$ is the only element at the first top left diagonal block of $\acute{\mathbf{\Theta}}$ in (49), thus $X$ is invertible if (49) holds. This is consistent with the fact that an invertible $W$ is implied by the matrix inequality in (57).

As a result, we have shown the equivalence between (33)–(34) and (47)–(48) for the case of $r_2 > r_1 > 0$. Meanwhile, it has been shown that (49) is equivalent to (57) which infers (35). Consequently, (33)–(35) are satisfied if (47)–(49) hold with some $W \in \mathbb{S}^n$ and $\{\alpha_i\}_{i=1}^{\widehat{3}+\kappa} \subset \mathbb{R}$. Thus it demonstrates that the existence of the feasible solutions of (47)–(49) ensures that the trivial solution $\boldsymbol{x}(t) \equiv \mathbf{0}_n$ of the closed-loop system in (19) with $\boldsymbol{w}(t) \equiv \mathbf{0}_q$ is uniformly asymptotically stable in $\mathbb{C}([-r, 0] ; \mathbb{R}^n)$ and (19) with (32) is dissipative.

Now for the case of $r_1 = r_2$, it is not difficult to show that a synthesis condition can be obtained by letting $d_2 = \delta_2 = 0$ in (47)–(49) with $\acute{Q}_2 = \acute{R}_2 = \acute{Y} = \mathsf{O}_n$ and $r_1 = r_2$, given the definition of $\widehat{3}$ and $\widehat{\mathrm{O}}$ in (24). The proof of such a synthesis condition for $r_1 = r_2$ follows the same procedures we have presented above with the substitutions $3 \leftarrow \widehat{3}$ and $4 \leftarrow \widehat{3} + 1$ and $d_2 = \delta_2 = 0$ in (51)–(63). Similarly, a synthesis condition for the case of $r_1 = 0$; $r_2 > 0$ can be obtained by letting $d_1 = \delta_1 = 0$ in (47)–(49) with the substitutions the substitutions $3 \leftarrow \widehat{3}$ and $4 \leftarrow \widehat{3} + 1$ and $\acute{Q}_1 = \acute{R}_1 = \mathsf{O}_n$ and $r_1 = 0$; $r_2 > 0$. ■

**Remark 6.** Note that Theorem 2 is specifically derived to solve a synthesis problem for (19). If an open-loop system is considered with $B_1 = \widetilde{B}_2(\tau) = \mathsf{O}_{n,p}$ and $B_4 = B_5(\tau) = \mathsf{O}_{m,p}$, then Theorem 1 should be applied instead of Theorem 2. This is because Theorem 2 is more conservative compared to Theorem 1 for a specific problem of stability analysis.

**Remark 7.** For $\{\alpha_i\}_{i=1}^{\widehat{3}+\kappa} \subset \mathbb{R}$ in (49), some values of $\alpha_i$ can have more significant impact on the feasibility of (49). For example, the value of $\alpha_{\widehat{3}}$ may have a significant impact on the feasibility of (49) since it may determine the feasibility of the very diagonal block related to $A_1$ in (49). A simple assignment for $\{\alpha_i\}_{i=1}^{\widehat{3}+\kappa} \subset \mathbb{R}$ can be $\alpha_i = 0$ for $i = 1 \cdots \widehat{3} + \kappa$ with $i \neq \widehat{3}$ which allows one to only adjust the value of $\alpha_{\widehat{3}}$ to use Theorem 2.

### 3.3. *An inner convex approximation solution of Theorem 1*

For a dissipative synthesis problem, Theorem 2 provides a convex solution. Nevertheless, the simplification in (56) can render Theorem 2 more conservative than Theorem 1, while the BMI in Theorem 1 cannot be solved by standard SDP solvers. To tackle these problems, an iterative algorithm is derived in this subsection based on the method proposed in Dinh *et al.* (2012). The algorithm provides an inner convex approximation solution for the BMI in (35), which can be initiated by a feasible solution of Theorem 2. Thus the advantage of both Theorem 1 and 2 are combined together in the proposed algorithm without the need to solve nonlinear optimization constraints.

First of all, note that (33) and (34) remain convex even when a synthesis problem is considered. Now it is obvious that (35) can be rewritten into

$$\mathfrak{U}(\mathbf{H}, K) := \mathsf{Sy}\left[\mathbf{P}^\top \mathbf{\Pi}\right] + \mathbf{\Phi} = \mathsf{Sy}\left(\mathbf{P}^\top \mathbf{B}\left[\left(I_{\widehat{3}+\kappa} \otimes K\right) \oplus \mathsf{O}_{p+m}\right]\right) + \widehat{\mathbf{\Phi}} \prec 0 \tag{64}$$

with $\mathbf{B} := \begin{bmatrix}\mathbf{B}_1 & \mathsf{O}_{n,m}\end{bmatrix}$ and $\widehat{\mathbf{\Phi}} := \mathsf{Sy}\left(\mathbf{P}^\top \begin{bmatrix}\mathbf{A} & \mathsf{O}_{n,m}\end{bmatrix}\right) + \mathbf{\Phi}$, where $\mathbf{P}$ is given in (40), and $\mathbf{A}$ and $\mathbf{B}_1$ are given in (20)–(21), and $\mathbf{H} := \begin{bmatrix}P_1 & P_2\end{bmatrix}$ with $P_1$ and $P_2$ in Theorem 1. It is important to stress here that $\widehat{\mathbf{\Phi}}$ is convex with respect to all the decision variables it contains. Considering the conclusions of Example 3 in Dinh *et al.* (2012), one can conclude that the function $\Delta\left(\bullet, \widetilde{\mathbf{G}}, \bullet, \widetilde{\mathbf{\Gamma}}\right)$, which is defined as

$$\begin{aligned}\Delta\left(\mathbf{G}, \widetilde{\mathbf{G}}, \mathbf{\Gamma}, \widetilde{\mathbf{\Gamma}}\right) := \begin{bmatrix}\mathbf{G}^\top - \widetilde{\mathbf{G}}^\top & \mathbf{\Gamma}^\top - \widetilde{\mathbf{\Gamma}}^\top\end{bmatrix}\left[Z \oplus (I_n - Z)\right]^{-1}\begin{bmatrix}\mathbf{G} - \widetilde{\mathbf{G}} \\ \mathbf{\Gamma} - \widetilde{\mathbf{\Gamma}}\end{bmatrix} \\ + \mathsf{Sy}\left(\widetilde{\mathbf{G}}^\top \mathbf{\Gamma} + \mathbf{G}^\top \widetilde{\mathbf{\Gamma}} - \widetilde{\mathbf{G}}^\top \widetilde{\mathbf{\Gamma}}\right) + \mathbf{T}\end{aligned} \tag{65}$$

with $Z \oplus (I_n - Z) \succ 0$ satisfying

$$\forall \mathbf{G}; \widetilde{\mathbf{G}} \in \mathbb{R}^{n\times l},\ \forall \mathbf{\Gamma}; \widetilde{\mathbf{\Gamma}} \in \mathbb{R}^{n\times l},\ \ \mathbf{T} + \mathsf{Sy}\left(\mathbf{G}^\top \mathbf{\Gamma}\right) \preceq \Delta\left(\mathbf{G}, \widetilde{\mathbf{G}}, \mathbf{\Gamma}, \widetilde{\mathbf{\Gamma}}\right),\ \ \mathbf{T} + \mathsf{Sy}\left(\mathbf{G}^\top \mathbf{\Gamma}\right) = \Delta(\mathbf{G}, \mathbf{G}, \mathbf{\Gamma}, \mathbf{\Gamma}), \tag{66}$$

is a psd-convex overestimate of $\acute{\Delta}(\mathbf{G}, \mathbf{\Gamma}) = \mathbf{T} + \mathsf{Sy}\left[\mathbf{G}^\top \mathbf{\Gamma}\right]$ with respect to the parameterization

$$\begin{bmatrix}\mathbf{vec}(\widetilde{\mathbf{G}}) \\ \mathbf{vec}(\widetilde{\mathbf{\Gamma}})\end{bmatrix} = \begin{bmatrix}\mathbf{vec}(\mathbf{G}) \\ \mathbf{vec}(\mathbf{\Gamma})\end{bmatrix}. \tag{67}$$

Let

$$\begin{gathered}\mathbf{T} = \widehat{\mathbf{\Phi}},\ \ \mathbf{G} = \mathbf{P} = \begin{bmatrix}\widehat{\mathsf{O}}_{n,n} & P_1 & P_2\widehat{I} & \mathsf{O}_{n,q} & \mathsf{O}_{n,m}\end{bmatrix}, \\ \widetilde{\mathbf{G}} = \widetilde{\mathbf{P}} = \begin{bmatrix}\widehat{\mathsf{O}}_{n,n} & \widetilde{P}_1 & \widetilde{P}_2\widehat{I} & \mathsf{O}_{n,q} & \mathsf{O}_{n,m}\end{bmatrix}, \\ \mathbf{H} = \begin{bmatrix}P_1 & P_2\end{bmatrix},\ \widetilde{\mathbf{H}} := \begin{bmatrix}\widetilde{P}_1 & \widetilde{P}_2\end{bmatrix},\ \widetilde{P}_1 \in \mathbb{S}^n,\ \widetilde{P}_2 \in \mathbb{R}^{n\times dn} \\ \mathbf{\Gamma} = \mathbf{B}\mathbf{K},\ \ \mathbf{K} = \left[\left(I_{\widehat{3}+\kappa} \otimes K\right) \oplus \mathsf{O}_{p+m}\right],\ \ \ \widetilde{\mathbf{\Gamma}} = \mathbf{B}\widetilde{\mathbf{K}},\ \ \widetilde{\mathbf{K}} = \left[\left(I_{\widehat{3}+\kappa} \otimes \widetilde{K}\right) \oplus \mathsf{O}_{p+m}\right]\end{gathered} \tag{68}$$

in (65) with $l = \widehat{3}n + \kappa n + q + m$ and $Z \oplus (I_n - Z) \succ 0$ and $\widehat{\mathbf{\Phi}}$, $\mathbf{H}$ and $K$ in line with the definition in (64), one can conclude that

$$\begin{aligned}\mathfrak{U}(\mathbf{H}, K) = \widehat{\mathbf{\Phi}} + \mathsf{Sy}\left[\mathbf{P}^\top \mathbf{B}\left[\left(I_{\widehat{3}+\kappa} \otimes K\right) \oplus \mathsf{O}_{p+m}\right]\right] \preceq \mathcal{S}\left(\mathbf{H}, \widetilde{\mathbf{H}}, K, \widetilde{K}\right) \\ := \widehat{\mathbf{\Phi}} + \mathsf{Sy}\left(\widetilde{\mathbf{P}}^\top \mathbf{B}\mathbf{K} + \mathbf{P}^\top \mathbf{B}\widetilde{\mathbf{K}} - \widetilde{\mathbf{P}}^\top \mathbf{B}\widetilde{\mathbf{K}}\right) + \begin{bmatrix}\mathbf{P}^\top - \widetilde{\mathbf{P}}^\top & \mathbf{K}^\top \mathbf{B}^\top - \widetilde{\mathbf{K}}^\top \mathbf{B}^\top\end{bmatrix}\left[Z \oplus (I_n - Z)\right]^{-1}[*]\end{aligned} \tag{69}$$

by (66), where $\mathcal{S}(\bullet, \widetilde{\mathbf{H}}, \bullet, \widetilde{K})$ in (69) is a psd-convex overestimate of $\mathfrak{U}(\mathbf{H}, K)$ in (64) with respect to the parameterization

$$\begin{bmatrix}\mathbf{vec}(\widetilde{\mathbf{H}}) \\ \mathbf{vec}(\widetilde{K})\end{bmatrix} = \begin{bmatrix}\mathbf{vec}(\mathbf{H}) \\ \mathbf{vec}(K)\end{bmatrix}. \tag{70}$$

From (69), it is obvious that $\mathcal{S}\left(\mathbf{H}, \widetilde{\mathbf{H}}, K, \widetilde{K}\right) \prec 0$ infers (64). Moreover, it is also true that $\mathcal{S}\left(\mathbf{H}, \widetilde{\mathbf{H}}, K, \widetilde{K}\right) \prec 0$ in (69) holds if and only if

$$\begin{bmatrix} \widehat{\mathbf{\Phi}} + \mathbf{Sy}\left(\widetilde{\mathbf{P}}^\top \mathbf{B}\mathbf{K} + \mathbf{P}^\top \mathbf{B}\widetilde{\mathbf{K}} - \widetilde{\mathbf{P}}^\top \mathbf{B}\widetilde{\mathbf{K}}\right) & \mathbf{P}^\top - \widetilde{\mathbf{P}}^\top & \mathbf{K}^\top \mathbf{B}^\top - \widetilde{\mathbf{K}}^\top \mathbf{B}^\top \\ * & -Z & \mathsf{O}_n \\ * & * & Z - I_n \end{bmatrix} \prec 0 \tag{71}$$

holds based on the application of the Schur complement given $Z \oplus (I_n - Z) \succ 0$. Now (64) is inferred by (71) which can be solved by standard numerical solvers of SDPs provided that the values of $\widetilde{\mathbf{H}}$ and $\widetilde{K}$ are known.

By compiling all the aforementioned procedures according to the expositions in Dinh *et al.* (2012), an iterative algorithm is constructed in Algorithm 1 where $\mathbf{x}$ consists of all the variables in $P_3$, $Q_1$, $Q_2$ $R_1$, $R_2$, $Y$ in Theorem 1 and $Z$ in (71). Furthermore, $\mathbf{H}$, $\widetilde{\mathbf{H}}$, $\mathbf{K}$ and $\widetilde{\mathbf{K}}$ in Algorithm 1 are defined in (68) and $\rho_1$, $\rho_2$ and $\varepsilon$ are given constants for regularizations and setting up error tolerance, respectively.

Based on the results in Dinh *et al.* (2012), one has to obtain certain initial data for $\widetilde{\mathbf{H}}$ and $\widetilde{K}$ to initialize Algorithm 1, which can be part of a feasible solution of (33)–(35) in Theorem 1. As a result, $\widetilde{P}_1 \leftarrow P_1$, $\widetilde{P}_2 \leftarrow P_2$ and $\widetilde{K} \leftarrow K$ is used for the initial data of $\widetilde{\mathbf{H}}$ and $\widetilde{K}$ in Algorithm 1 if $P_1$, $P_2$ and $K$ are a feasible solutions of (33)–(35). Generally speaking, acquiring a feasible solution of Theorem 1 may not be an easy task. Nevertheless, as what has been proposed in Theorem 2, initial values of $\widetilde{P}_1$, $\widetilde{P}_2$ and $\widetilde{K}$ can be supplied by solving the constraints in (47)–(49) with given values[6] of $\{\alpha_i\}_{i=1}^{\widehat{3}+\kappa}$.

---

**Algorithm 1:** An inner convex approximation solution for Theorem 1 with $r_2 > r_1 > 0$

---

**begin**

  **solve** Theorem 2 with given $\alpha_i$ to obtain a feasible $K$, and then **solve** Theorem 1 with the previous $K$ to obtain $\mathbf{H} = \begin{bmatrix} P_1 & P_2 \end{bmatrix}$.

  **update** $\widetilde{\mathbf{H}} \longleftarrow \mathbf{H}$, $\widetilde{K} \longleftarrow K$,

  **solve** $\min\limits_{\mathbf{x},\mathbf{H},K} \operatorname{tr}\left[\rho_1[*]\left(\mathbf{H} - \widetilde{\mathbf{H}}\right)\right] + \operatorname{tr}\left[\rho_2[*]\left(K - \widetilde{K}\right)\right]$ subject to (33)–(34) and (71) to obtain $\mathbf{H}$ and $K$

  **while** $\dfrac{\left\|\begin{bmatrix} \mathbf{vec}(\mathbf{H}) \\ \mathbf{vec}(K) \end{bmatrix} - \begin{bmatrix} \mathbf{vec}(\widetilde{\mathbf{H}}) \\ \mathbf{vec}(\widetilde{K}) \end{bmatrix}\right\|_\infty}{\left\|\begin{bmatrix} \mathbf{vec}(\widetilde{\mathbf{H}}) \\ \mathbf{vec}(\widetilde{K}) \end{bmatrix}\right\|_\infty + 1} \geq \varepsilon$ **do**

    **update** $\widetilde{\mathbf{H}} \longleftarrow \mathbf{H}$, $\widetilde{K} \longleftarrow K$;

    **solve** $\min\limits_{\mathbf{x},\mathbf{H},K} \operatorname{tr}\left[\rho_1[*]\left(\mathbf{H} - \widetilde{\mathbf{H}}\right)\right] + \operatorname{tr}\left[\rho_2[*]\left(K - \widetilde{K}\right)\right]$ subject to (33)–(34) and (71) to obtain $\mathbf{H}$ and $K$;

  **end**

**end**

---

[6]Note that as we have elaborated in Remark 7 that one may apply Theorem 2 with $\alpha_i = 0$ for $i = 1 \cdots \widehat{3} + \kappa, i \neq \widehat{3}$ which allow users to only adjust the value of $\alpha_{\widehat{3}}$ to solve the conditions in Theorem 2

**Remark 8.** If a convex objective function is considered in Theorem 1, for instance $\mathbb{L}^2$ gain $\gamma > 0$ minimization, a termination criterion Dinh *et al.* (2012) can be added to Algorithm 1 in order to characterize the progress of the objective function between each adjacent iteration. Nonetheless, such a condition has not been concerned by the tests of our numerical examples in this paper.

**Remark 9.** For the delay values $r_2 > 0$; $r_1 = 0$ or $r_2 = r_1 > 0$, Algorithm 1 can be utilized via the corresponding synthesis conditions with appropriate parameter assignments as stated in the statements of Theorem 1 and 2.

Since we have proposed many technical results in this paper, a summary concerning their relations is presented as follows:

- The first important technical result is the decomposition scenario in Proposition 1. This enables us to denote general distributed delays in terms of the products between constants and some appropriate functions.
- By using Proposition 1, one can derive the synthesis results in Theorem 1 where the synthesis condition is characterized by optimization constraints of finite dimensions thanks to the application of the integral inequality proposed in (B.5).
- Theorem 2 has been proposed as a convexification of the BMI in Theorem 1 via the application of the Projection Lemma.
- Algorithm 1 has been further proposed to solve the BMI in Theorem 1 based on the inner convex approximation algorithm. The initial value of Algorithm 1 can be provided by solving the synthesis condition in Theorem 2.

## 4. Numerical examples

In this section, two numerical examples are presented to demonstrate the effectiveness of our proposed methodologies. The numerical tests are conducted in Matlab environment using Yalmip Löfberg (2004) as the optimization interface. Moreover, we use SDPT3 Toh *et al.* (2012) for solving SDPs numerically.

### *4.1. Stability and dissipative analysis of a linear system with a time-varying distributed delay*

Consider a system of the form (1) with any $r(\cdot) \in \mathbb{M}\big(\mathbb{R} ; [r_1, r_2]\big)$ and the state space matrices

$$
\begin{aligned}
&A_1 = \begin{bmatrix} 0.1 & 0 \\ 0 & -1 \end{bmatrix},\ \widetilde{A}_2(\tau) = \begin{bmatrix} 0.3\mathrm{e}^{\cos(5\tau)} - 0.1\mathrm{e}^{\sin(5\tau)} - 0.4 & 0.01\mathrm{e}^{\cos(5\tau)} - 0.1\mathrm{e}^{\sin(5\tau)} + 1 \\ \ln(2-\tau) - 1 & 0.4 - 0.3\mathrm{e}^{\cos(5\tau)} \end{bmatrix}, \\
&B_1 = \widetilde{B}_2(\tau) = B_4 = \widetilde{B}_5(\tau) = \begin{bmatrix} 0 \\ 0 \end{bmatrix}, D_1 = \begin{bmatrix} 0.1 \\ 0.2 \end{bmatrix}, C_1 = \begin{bmatrix} -0.1 & 0.2 \\ 0 & 0.1 \end{bmatrix}, \\
&\widetilde{C}_2(\tau) = \begin{bmatrix} 0.2\mathrm{e}^{\sin(5\tau)} - 0.11 & 0.1 - 0.5\ln(2-\tau) \\ 0.1\mathrm{e}^{\sin(5\tau)} & 0.14\mathrm{e}^{\cos(5\tau)} - 0.2\mathrm{e}^{\sin(5\tau)} \end{bmatrix},\ D_2 = \begin{bmatrix} 0.12 \\ 0.1 \end{bmatrix}.
\end{aligned}
\tag{72}
$$

Moreover, let

$$
J_1 = -\gamma I_m,\ \ \widetilde{J} = I_m,\ \ J_2 = \mathsf{O}_{m,q},\ \ J_3 = \gamma I_q \tag{73}
$$

for the supply rate function in (32) where the objective is to calculate the minimum value of $\mathbb{L}^2$ gain $\gamma$. Note that all the controller gains in (72) are of zero values, and the distributed delays in (72) contain different types of functions.

To the best of our knowledge, no existing approaches, neither time nor frequency-domain based methods, can analyze the stability of (1) with the parameters in (72). Note that since $r(t)$ is time-varying and its expression is unknown, hence the distributed delay kernels in (72) may not be approximated over $[-r(t), 0]$ via the approaches in Münz *et al.* (2009); Seuret *et al.* (2015). For the same reason, the distributed delays may not be easily analyzed in frequency domain analytically via the existing methods in Kharitonov *et al.* (2009); Breda *et al.* (2015); Vyhlídal & Zítek (2014). Finally, no existing methods may calculate the $\mathbb{L}^2$ gain of the system considered in this subsection.

By observing the functions inside of $\widetilde{A}_2(\cdot), \widetilde{C}_2(\cdot)$ in (72), we choose

$$\begin{gathered}
\boldsymbol{f}_1(\tau) = \boldsymbol{f}_2(\tau) = \begin{bmatrix} 1 \\ \mathrm{e}^{\sin(5\tau)} \\ \mathrm{e}^{\cos(5\tau)} \\ \ln(2-\tau) \end{bmatrix}, \quad \boldsymbol{\varphi}_1(\tau) = \boldsymbol{\varphi}_2(\tau) = \begin{bmatrix} \cos(5\tau)\mathrm{e}^{\sin(5\tau)} \\ \sin(5\tau)\mathrm{e}^{\cos(5\tau)} \\ \dfrac{1}{\tau-2} \end{bmatrix}, \\
M_1 = M_2 = \begin{bmatrix} 0 & 0 & 0 & 0 & 0 & 0 & 0 \\ 5 & 0 & 0 & 0 & 0 & 0 & 0 \\ 0 & -5 & 0 & 0 & 0 & 0 & 0 \\ 0 & 0 & 1 & 0 & 0 & 0 & 0 \end{bmatrix}
\end{gathered} \tag{74}$$

for the functions $\boldsymbol{f}_1(\cdot), \boldsymbol{f}_2(\cdot)$ and $\boldsymbol{\varphi}_1(\cdot), \boldsymbol{\varphi}_2(\cdot)$ in Proposition 1, which corresponds to $d_1 = d_2 = 4$, $\delta_1 = \delta_2 = 3$, $n = m = 2$, $q = 1$, and

$$\begin{aligned}
A_2 = A_3 &= \begin{bmatrix} 0 & 0 & 0 & 0 & 0 & 0 & -0.4 & 1 & -0.1 & -0.1 & 0.3 & 0.01 & 0 & 0 \\ 0 & 0 & 0 & 0 & 0 & 0 & -1 & 0.4 & 0 & 0 & 0 & -0.3 & 1 & 0 \end{bmatrix}, \quad B_2 = B_3 = \mathsf{O}_{2\times 7} \\
C_2 = C_3 &= \begin{bmatrix} 0 & 0 & 0 & 0 & 0 & 0 & -0.11 & 0.1 & 0.2 & 0 & 0 & 0 & 0 & -0.5 \\ 0 & 0 & 0 & 0 & 0 & 0 & 0 & 0 & 0.1 & -0.2 & 0 & 0.14 & 0 & 0 \end{bmatrix}, \quad B_5 = B_6 = \mathsf{O}_{2\times 7}.
\end{aligned} \tag{75}$$

Now apply Theorem 1 to (19) with the parameters in (72)–(75), where the conditions in Theorem 1 are all convex in this case. It produces the results in Tables 1–2, where several detectable delay boundaries are presented with the corresponding $\min\gamma$.

| $[r_1, r_2]$ | $[0.98, 1.25]$ | $[1, 1.23]$ | $[1.02, 1.21]$ | $[1.04, 1.19]$ |
|---|---|---|---|---|
| $r_3 = r_2 - r_1$ | 0.27 | 0.23 | 0.19 | 0.15 |
| $\min\gamma$ | 0.5511 | 0.51356 | 0.48277 | 0.45692 |

**Table 1:** $\min\gamma$ produced with decreasing values of $r_3$

| $[r_1, r_2]$ | $[0.8, 1.07]$ | $[1, 1.27]$ | $[1.2, 1.47]$ | $[1.32, 1.59]$ |
|---|---|---|---|---|
| $r_3 = r_2 - r_1$ | 0.27 | 0.27 | 0.27 | 0.27 |
| $\min\gamma$ | 0.35556 | 0.59179 | 1.7935 | 25.9774 |

**Table 2:** $\min\gamma$ produced with a fixed value for $r_3$

The results of $\min\gamma$ in Table 1 indicate that smaller $r_3$ can lead to smaller $\min\gamma$ values. Indeed, it is more difficult to make the system to be dissipative for all $r(\cdot) \in \mathbb{M}(\mathbb{R} ⨾ [r_1, r_2])$ with a large value of $r_3$ than for all $r(\cdot) \in \mathbb{M}(\mathbb{R} ⨾ [\acute{r}_1, \acute{r}_2])$ with a smaller value of $\acute{r}_3 = \acute{r}_2 - \acute{r}_1$ if $[\acute{r}_1, \acute{r}_2] \subset [r_1, r_2]$. On the other hand, the values of $\min\gamma$ in table 2 show that the values of $r_1$ and $r_2$ can significantly affect the resulting $\min\gamma$ even with a fixed $r_3 = r_2 - r_1$.

In order to partially verify the results in Tables 1 and 2, we utilize the frequency domain method in Breda *et al.* (2015) to (72) assuming that $r(\cdot) \in \mathbb{M}(\mathbb{R} ⨾ [r_1, r_2])$ is an unknown function with a constant value. (Note that an unknown $r(\cdot)$ with a constant value is an option for $r(\cdot) \in \mathbb{M}(\mathbb{R} ⨾ [r_1, r_2])$) The result shows that the system with a constant value of $r$ is stable over $[0.61, 1.64]$, which is consistent with the results in Tables 1 and 2. This is because the results in Tables 1 and 2 infer that the system with a constant delay value is stable over the intervals therein. which are all the subsets of $[0.61, 1.64]$.

**Remark 10.** Note that the values of $\min\gamma$ in Tables 1–2 are valid for any $r(\cdot) \in \mathbb{M}(\mathbb{R} ⨾ [r_1, r_2])$ with given $r_1$ and $r_2$ since the proposed methods in this paper guarantee that the system with (73) is dissipative for any $r(\cdot) \in \mathbb{M}(\mathbb{R} ⨾ [r_1, r_2])$. This is also true for other options for dissipative constraints.

*4.2. Dissipative stabilization of a linear system with a time-varying distributed delay*

Consider a system of the form (1) with any $r(\cdot) \in \mathbb{M}(\mathbb{R} ⨾ [0.5, 1])$ and the state space parameters

$$
\begin{aligned}
&A_1 = \begin{bmatrix} -1 & -1.9 \\ 0 & 0.1 \end{bmatrix}, \ \widetilde{A}_2(\tau) = \begin{bmatrix} 0.2\cos(\mathrm{e}^{\tau}) + 0.1\sin(\mathrm{e}^{\tau}) & 0.01\cos(\mathrm{e}^{\tau}) - 0.1\sin(\mathrm{e}^{\tau}) \\ 0 & -0.4\cos(\mathrm{e}^{\tau}) \end{bmatrix}, \ \tau \in [-r_1, 0] \\
&\widetilde{A}_2(\tau) = \begin{bmatrix} 0.2\cos(\mathrm{e}^{\tau}) + 0.1\sin(\mathrm{e}^{\tau}) - 0.2 & 0.01\cos(\mathrm{e}^{\tau}) - 0.1\sin(\mathrm{e}^{\tau}) + 1 \\ \ln(2 - \cos(\tau)) - 1.2 & 1 - 0.4\cos(\mathrm{e}^{\tau}) \end{bmatrix}, \ \ \tau \in [-r(t), -r_1] \\
&B_1 = \begin{bmatrix} 0 \\ 1 \end{bmatrix}, \ \ \widetilde{B}_2(\tau) = \begin{bmatrix} 0.1\sin(\mathrm{e}^{\tau}) - 0.1 \\ 0.12\cos(\mathrm{e}^{\tau}) + 0.1 \end{bmatrix}, D_1 = \begin{bmatrix} 0.01 \\ 0.02 \end{bmatrix}, C_1 = \begin{bmatrix} 0.1 & 0.15 \\ 0 & -0.2 \end{bmatrix}, \\
&\widetilde{C}_2(\tau) = \begin{bmatrix} 0.2\sin(\mathrm{e}^{\tau}) + 0.1 & 0.1 \\ -0.2\sin(\mathrm{e}^{\tau}) & 0.3\sin(\mathrm{e}^{\tau}) - 0.1\cos(\mathrm{e}^{\tau}) \end{bmatrix}, \ \ B_4 = \begin{bmatrix} 0 \\ 0.1 \end{bmatrix} \\
&\widetilde{B}_5(\tau) = \begin{bmatrix} 0 \\ 0.1 - 0.1\sin(\mathrm{e}^{\tau}) \end{bmatrix}, \ \ D_2 = \begin{bmatrix} 0.1 \\ 0.2 \end{bmatrix}.
\end{aligned}
\tag{76}
$$

Moreover, let

$$
J_1 = -\gamma I_m, \ \ \widetilde{J} = I_m, \ \ J_2 = \mathsf{O}_{m,q}, \ \ J_3 = \gamma I_q \tag{77}
$$

for the supply rate function in (32) to calculate the minimum value of $\mathbb{L}^2$ gain $\gamma$.

According to our best knowledge, no existing methods can find a controller for (1) with the parameters in (76).

By observing the functions inside of $\widetilde{A}_2(\cdot), \widetilde{B}_2(\cdot), \widetilde{C}_2(\cdot), \widetilde{B}_5(\cdot)$, we choose $\boldsymbol{f}_1(\cdot), \boldsymbol{f}_2(\cdot)$ and $\boldsymbol{\varphi}_1(\cdot), \boldsymbol{\varphi}_2(\cdot)$ in

Proposition 1 to be

$$
\begin{gathered}
\boldsymbol{f}_1(\tau) = \begin{bmatrix} 1 \\ \sin(\mathrm{e}^{\tau}) \\ \cos(\mathrm{e}^{\tau}) \end{bmatrix}, \quad \boldsymbol{f}_2(\tau) = \begin{bmatrix} 1 \\ \sin(\mathrm{e}^{\tau}) \\ \cos(\mathrm{e}^{\tau}) \\ \ln(2-\cos\tau) \end{bmatrix}, \quad \boldsymbol{\varphi}_1(\tau) = \begin{bmatrix} \mathrm{e}^{\tau}\cos(\mathrm{e}^{\tau}) \\ \mathrm{e}^{\tau}\sin(\mathrm{e}^{\tau}) \end{bmatrix}, \quad \boldsymbol{\varphi}_2(\tau) = \begin{bmatrix} \mathrm{e}^{\tau}\cos(\mathrm{e}^{\tau}) \\ \mathrm{e}^{\tau}\sin(\mathrm{e}^{\tau}) \\ \dfrac{\sin\tau}{2-\cos\tau} \end{bmatrix} \\
M_1 = \begin{bmatrix} 0 & 0 & 0 & 0 & 0 \\ 1 & 0 & 0 & 0 & 0 \\ 0 & -1 & 0 & 0 & 0 \end{bmatrix}, \quad M_2 = \begin{bmatrix} 0 & 0 & 0 & 0 & 0 & 0 & 0 \\ 1 & 0 & 0 & 0 & 0 & 0 & 0 \\ 0 & -1 & 0 & 0 & 0 & 0 & 0 \\ 0 & 0 & 1 & 0 & 0 & 0 & 0 \end{bmatrix}
\end{gathered} \tag{78}
$$

with $d_1 = 3, d_2 = 4$, $\delta_1 = 2, \delta_3 = 3$, $n = m = 2$, $q = 1$, and

$$
\begin{aligned}
A_2 &= \begin{bmatrix} 0 & 0 & 0 & 0 & 0 & 0 & 0.1 & -0.1 & 0.2 & 0.01 \\ 0 & 0 & 0 & 0 & 0 & 0 & 0 & 0 & 0 & -0.4 \end{bmatrix}, \quad A_3 = \begin{bmatrix} 0 & 0 & 0 & 0 & 0 & 0 & -0.2 & 1 & 0.1 & -0.1 & 0.2 & 0.01 & 0 & 0 \\ 0 & 0 & 0 & 0 & 0 & 0 & -1.2 & 1 & 0 & 0 & 0 & -0.4 & 1 & 0 \end{bmatrix} \\
B_2 &= \begin{bmatrix} 0 & 0 & -0.1 & 0.1 & 0 \\ 0 & 0 & 0.1 & 0 & 0.12 \end{bmatrix}, \quad B_3 = \begin{bmatrix} 0 & 0 & 0 & -0.1 & 0.1 & 0 & 0 \\ 0 & 0 & 0 & 0.1 & 0 & 0.12 & 0 \end{bmatrix} \\
C_2 &= \begin{bmatrix} 0 & 0 & 0 & 0 & 0.1 & 0.1 & 0.2 & 0 & 0 & 0 \\ 0 & 0 & 0 & 0 & 0 & 0 & -0.2 & 0.3 & 0 & -0.1 \end{bmatrix}, \quad C_3 = \begin{bmatrix} 0 & 0 & 0 & 0 & 0 & 0 & 0.1 & 0.1 & 0.2 & 0 & 0 & 0 & 0 & 0 \\ 0 & 0 & 0 & 0 & 0 & 0 & 0 & 0 & -0.2 & 0.3 & 0 & -0.1 & 0 & 0 \end{bmatrix} \\
B_5 &= \begin{bmatrix} 0 & 0 & 0 & 0 & 0 \\ 0 & 0 & 0.1 & -0.1 & 0 \end{bmatrix}, \quad B_6 = \begin{bmatrix} 0 & 0 & 0 & 0 & 0 & 0 & 0 \\ 0 & 0 & 0 & 0.1 & -0.1 & 0 & 0 \end{bmatrix}.
\end{aligned} \tag{79}
$$

Now apply Algorithm 1 to (19) with the parameters in (76)–(79) and with $\alpha_1 = \alpha_2 = \alpha_i = 0, i = 4 \cdots 12$ and $\alpha_3 = 0.5$ for the initialization of Algorithm 1 via Theorem 2. It produces the controller gains and the corresponding $\min\gamma$ in Tables 3, where NoIs stands for the number of iterations in the while loop inside of Algorithm 1.

| Controller gain $K$ | $\begin{bmatrix} 0.4182 \\ -2.7551 \end{bmatrix}^{\top}$ | $\begin{bmatrix} 0.5011 \\ -2.7108 \end{bmatrix}^{\top}$ | $\begin{bmatrix} 0.5787 \\ -2.6595 \end{bmatrix}^{\top}$ | $\begin{bmatrix} 0.6505 \\ -2.6021 \end{bmatrix}^{\top}$ |
|---|---|---|---|---|
| $\min\gamma$ | 0.36657 | 0.3607 | 0.3551 | 0.3498 |
| NoIs | 10 | 20 | 30 | 40 |

**Table 3:** Controller gains with $\min\gamma$ produced with different iterations with $a = 1$

Since $r(t)$ in this paper is time-varying and its expression is unknown, hence existing frequency-domain-based approaches may not be directly applied to analyze the stability of the resulting closed-loop systems obtained by our methods. To partially verify our synthesis results in Tables 3, we confine $r(t)$ to be an unknown constant $\widehat{r} \in [r_1, r_2]$. This allows one to apply the spectral method in Breda *et al.* (2015) to calculate the spectral abscissa of the spectrum of the resulting closed-loop systems with a constant delay. Since our synthesis results indicate that any resulting closed-loop system is stable for all $r(\cdot) \in \mathbb{M}\left(\mathbb{R} \,⨾\, [r_1, r_2]\right)$, thus the same closed-loop systems with a constant delay $\widehat{r}$ are stable for $\widehat{r} \in [r_1, r_2]$ as the case of $r(t) = \widehat{r}$ is included by $\mathbb{M}\left(\mathbb{R} \,⨾\, [r_1, r_2]\right)$. The numerical results produced by Breda *et al.* (2015) show that all the resulting closed-loop systems are stable for $\widehat{r} \in [r_1, r_2]$ with the assumption that $r(t) = \widehat{r}$ is a constant delay.

For numerical simulation, consider the closed-loop systems stabilized by the controller $K = \begin{bmatrix} 0.6505 & -2.6021 \end{bmatrix}$ in Table 3. Specifically, assume $t_0 = 0$, $\boldsymbol{z}(t) = \mathbf{0}_2, t < 0$, and $\boldsymbol{\phi}(\tau) = \begin{bmatrix} 50 & 30 \end{bmatrix}^\top, \tau \in [-1, 0]$ as the initial condition, and $\boldsymbol{w}(t) = 50 \sin 10t(\mathsf{u}(t) - \mathsf{u}(t-5))$ as the disturbance where $\mathsf{u}(t)$ is the Heaviside step function. Moreover, we consider a time-varying delay $r(t) = 0.75 + 0.25\cos(100t)$ which[7] exhibits strong oscillation. Numerical simulation is performed in Simulink with the aforementioned data via the ODE solver `ode8` with $0.0001$ as the fundamental sampling time. The result of our simulation is presented in Figures 1–3 concerning the trajectories of the states, outputs and the controller compensation of the closed-loop systems. Note that the update method of the Matlab function block in Simulink is set as 'discrete' for our simulation. Note that also the modeling of the distributed delays for simulation is attained by discretizing the integrals with the trapezoidal rule

$$\int_{-r_2}^{0} F(t,\tau)\boldsymbol{x}(t+\tau)\mathrm{d}\tau \approx \frac{r_2}{n}\left(\frac{F(t,-r_2)\boldsymbol{x}(t-r_2)}{2} + \sum_{k=1}^{n-1} F\left(t, \frac{kr_2}{n} - r_2\right)\boldsymbol{x}(t + \frac{kr_2}{n} - r_2) + \frac{F(t,0)\boldsymbol{x}(t)}{2}\right) \tag{80}$$

where

$$F(t,\tau) := \begin{cases} \widetilde{F}(\tau) & \forall \tau \in [-r(t), 0] \\ 0 & \forall \tau \in [-r_2, r(t)) \end{cases} \tag{81}$$

and $\widetilde{F}(\tau)$ is piecewise continuous on $[-r(t), 0]$.

**Remark 11.** Note that (81) enables one to discretize $\int_{-r(t)}^{0} \widetilde{F}(\tau)\boldsymbol{x}(t+\tau)\mathrm{d}\tau$ via (80) which avoids dealing with $\int_{-r(t)}^{0} \widetilde{F}(\tau)\boldsymbol{x}(t+\tau)\mathrm{d}\tau$ directly.

**Remark 12.** Due to the absence of proper numerical solvers in Simulink for delay systems, we can only use an ODE solver (ode8) in Simulink to conduct our simulation. Since we cannot predict the potential problems of using an ODE solver to a delay system, thus the numerical results in Figures 1–3 only give an estimation of the actual behavior of the system trajectories and output, and the numerical accuracy in this case may not be guaranteed.

**Remark 13.** The results in Figures 1–3 can clearly demonstrate the effectiveness of the proposed stabilization method considering a time-varying delay $r(t) = 0.75 + 0.25\cos(100t)$. Note that the abrupt change around $t = 5$ in Figures 3 is due to the form of the disturbance signal $\boldsymbol{w}(t) = 50\sin 10t(\mathsf{u}(t) - \mathsf{u}(t-5))$ which satisfies $\forall t > 5$, $\boldsymbol{w}(t) = 0$.

## 5. Conclusion

In this paper, new methods on the dissipative state feedback stabilization of a linear system with distributed delays (1) have been proposed, where the distributed delay kernels can be any $\mathbb{L}^2$ function and the time-varying delay function is bounded and measurable. The key step of deriving the synthesis condition in Theorem 1 is the application of the novel inequality proposed in Lemma 6 together with the decomposition scenario in Proposition 1, which results in LMIs with finite dimensions as explained in subsection 3.1. Though (35) in Theorem 1 is bilinear, it has been shown in Theorem 2 that convex conditions (47)–(49) can be constructed via the application of Projection Lemma to (35). Moreover, an iterative algorithm has been proposed in Algorithm 1 as an inner approximation solution to (35) in Theorem 1, which can be initiated through a feasible solution of Theorem 2. On the other hand, it is worthy of mentioning that our synthesis conditions can also handle the cases of $r_1 = r_2$ or $r_1 = 0$; $r_2 > 0$, based on the application of empty matrices. Finally, the proposed methodologies can handle any real-time application if they can be modeled by the general distributed delay system considered in this paper. This includes the cases where $r(t)$ is a stochastic and bounded function.

[7]Note that this function satisfies $\forall t \geq t_0, r_1 = 0.5 \leq r(t) \leq 1 = r_2$

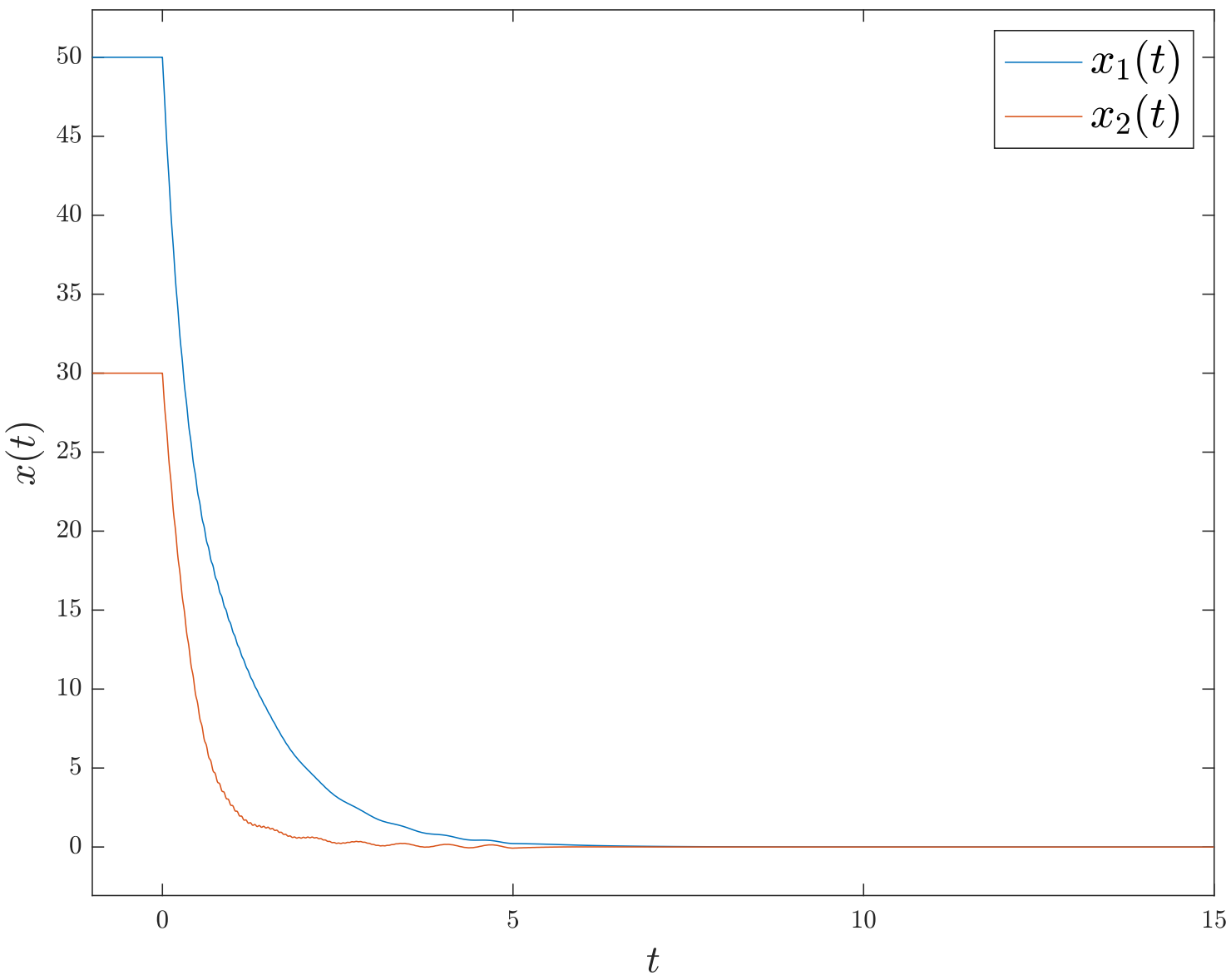

**Figure 1:** The close-loop system's trajectory $\boldsymbol{x}(t)$ with $K = \begin{bmatrix} 0.6505 & -2.6021 \end{bmatrix}$ in Table 3

### Acknowledgements

Qian Feng would like to pay tribute to Prof. Sing Kiong Nguang (the second author of this paper), who recently passed away. As the Ph.D. supervisor of Qian Feng, Prof. Nguang contributed great inspiration and guidance to him. Finally, though our proof for Theorem 4 did not ultimately employ the approach suggested by Prof. Keqin Gu, Qian Feng still would like to sincerely thank him for the fruitful discussion on how to prove a Krasovskiĭ stability theorem considering the Caratheodory conditions.

### Appendix A. Important Lemmas

The following properties of the Kronecker product will be used throughout this paper, which are derived from the definition of the Kronecker product and the property $(A \otimes B)(C \otimes D) = AC \otimes BD$.

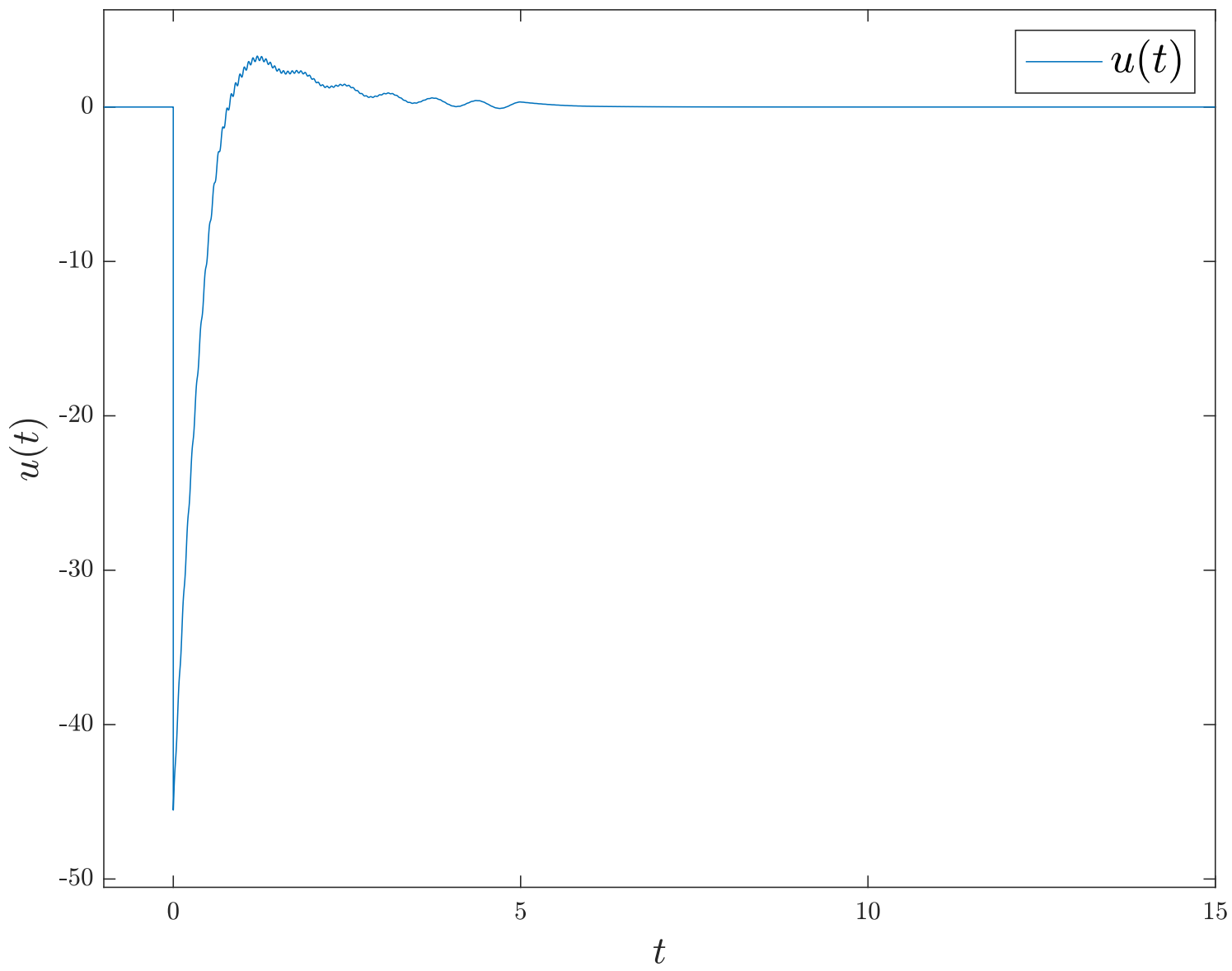

**Figure 2:** The trajectory of the controller effort $\boldsymbol{u}(t) = K\boldsymbol{x}(t)$ with $K = \begin{bmatrix} 0.6505 & -2.6021 \end{bmatrix}$ in Table 3

**Lemma 2.** $\forall X \in \mathbb{R}^{n\times m}, \ \forall Y \in \mathbb{R}^{m\times p}, \ \ \forall Z \in \mathbb{R}^{q\times r},$

$$(X \otimes I_q)(Y \otimes Z) = (XY) \otimes (I_q Z) = (XY) \otimes Z = (XY) \otimes (ZI_r) = (X \otimes Z)(Y \otimes I_r). \tag{A.1}$$

*Moreover,* $\forall X \in \mathbb{R}^{n\times m}$, *we have*

$$\begin{bmatrix} A & B \\ C & D \end{bmatrix} \otimes X = \begin{bmatrix} A\otimes X & B\otimes X \\ C\otimes X & D\otimes X \end{bmatrix} \tag{A.2}$$

*for any* $A, B, C, D$ *with appropriate dimensions.*

The following property of the commutation matrix Magnus & Neudecker (1979) are utilized throughout this paper.

**Lemma 3.**

$$\begin{aligned} \forall X \in \mathbb{R}^{d\times \delta}, \ \ \forall Y \in \mathbb{R}^{n\times m} \ \ \mathsf{K}_{(n,d)} \left( X\otimes Y\right) \mathsf{K}_{(\delta,m)} = Y \otimes X \\ \forall m,n \in \mathbb{N}, \quad \mathsf{K}^{-1}_{(n,m)} = \mathsf{K}_{(m,n)} = \mathsf{K}^{\top}_{(n,m)} \end{aligned} \tag{A.3}$$

*where* $\mathsf{K}_{(n,d)}$ *is the commutation matrix defined by the identity*

$$\forall A \in \mathbb{R}^{n\times d}, \ \ \mathsf{K}_{(n,d)}\, \mathbf{vec}\left(A\right) = \mathbf{vec}\left(A^{\top}\right)$$

*which follows the definition in Magnus & Neudecker (1979), where* $\mathbf{vec}(\cdot)$ *stands for the vectorization of a matrix. See Section 4.2 of Dhrymes (2013) for the definition and more details of* $\mathbf{vec}(\cdot)$.

**Remark 14.** Note that for $\mathsf{K}_{(n,d)}$, we have $\mathsf{K}_{(n,1)} = \mathsf{K}_{(1,n)} = I_n, \forall n \in \mathbb{N}$ which gives the identity

$$\mathsf{K}_{(n,d)} \left( \boldsymbol{f}(\tau) \otimes I_n \right) = \mathsf{K}_{(n,d)} \left( \boldsymbol{f}(\tau) \otimes I_n \right) K_{(1,n)} = I_n \otimes \boldsymbol{f}(\tau) \tag{A.4}$$

with $\boldsymbol{f}(\tau) \in \mathbb{R}^d$. The commutation matrix $\mathsf{K}_{(n,d)}$ can be numerically implemented by $\mathsf{K}_{(n,d)} =$ `vecperm(d, n)` in Matlab where `vecperm` is a function in The Matrix Computation Toolbox for MATLAB Higham (2002).

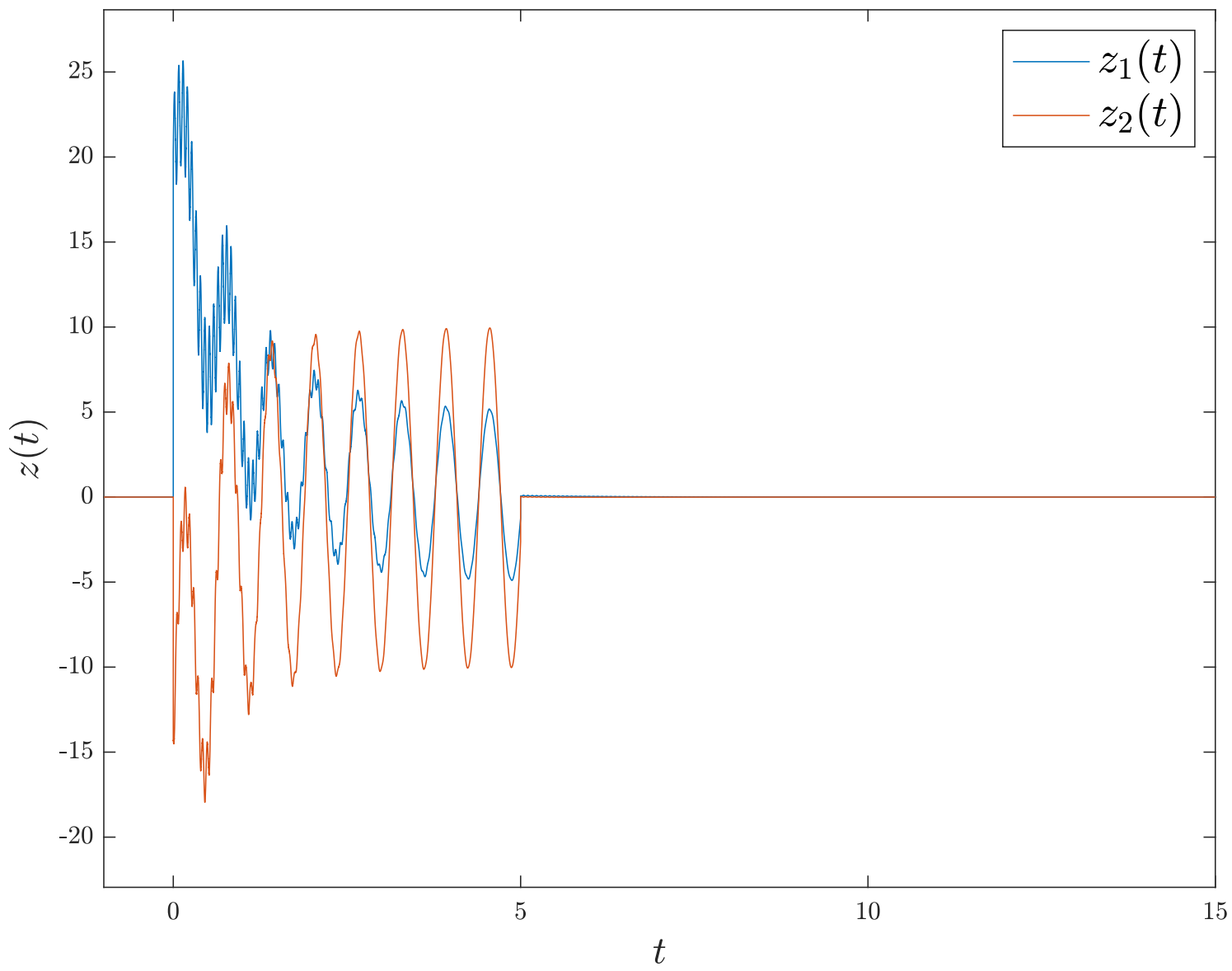

**Figure 3:** The output of the closed-loop system $\boldsymbol{z}(t)$ with $K = \begin{bmatrix} 0.6505 & -2.6021 \end{bmatrix}$ in Table 3

**Lemma 4.** *Consider the functional differential equation*

$$
\begin{aligned}
&\widetilde{\forall} t \geq t_0, \quad \dot{\boldsymbol{x}}(t) = \boldsymbol{f}(t, \mathbf{x}_t(\cdot)), \\
&\forall \theta \in [-r, 0], \quad \boldsymbol{x}(t_0 + \theta) = \mathbf{x}_{t_0}(\theta) = \boldsymbol{\phi}(\theta), \quad r > 0 \\
&\forall t \in \mathbb{R}, \ \mathbf{0}_n = \boldsymbol{f}(t, \mathbf{0}_n(\cdot))
\end{aligned} \tag{A.5}
$$

*where* $t_0 \in \mathbb{R}$ *and* $\boldsymbol{f} : \mathbb{R} \times \mathbb{C}\left([-r, 0] ; \mathbb{R}^n\right) \to \mathbb{R}^n$ *satisfies the Caratheodory conditions in section 2.6 of Hale & Lunel (1993) and*

$$
\exists c(\cdot) \in \mathbb{R}_{>0}^{\mathbb{R}_{>0}}, \ \forall \delta > 0, \ \forall \boldsymbol{\phi}(\cdot) \in \mathbb{C}_{\delta}\left([-r, 0] ; \mathbb{R}^n\right), \ \widetilde{\forall} t \in \mathbb{R}, \quad \left\|\boldsymbol{f}\left(t, \boldsymbol{\phi}(\cdot)\right)\right\|_1 < c(\delta). \tag{A.6}
$$

*Then the trivial solution* $\boldsymbol{x}(t) \equiv \mathbf{0}_n$ *of* (A.5) *is uniformly asymptotically stable in* $\mathbb{C}([-r, 0] ; \mathbb{R}^n)$ *if there exist* $\alpha_1(\cdot); \alpha_2(\cdot); \alpha_3(\cdot) \in \mathcal{K}_{\infty}$, *and a continuous functional* $\mathsf{v} : \mathbb{R} \times \mathbb{C}([-r, 0] ; \mathbb{R}^n) \to \mathbb{R}$ *with* $\forall t \in \mathbb{R}$, $\mathsf{v}(t, \mathbf{0}_n(\cdot)) = 0$ *such that*

$$
\forall t \in \mathbb{R}, \ \forall \boldsymbol{\phi}(\cdot) \in \mathbb{C}([-r_2, 0] ; \mathbb{R}^n), \ \alpha_1\left(\|\boldsymbol{\phi}(0)\|_2\right) \leq \mathsf{v}(t, \boldsymbol{\phi}(\cdot)) \leq \alpha_2\left(\|\boldsymbol{\phi}(\cdot)\|_{\infty}\right), \tag{A.7}
$$

$$
\widetilde{\forall} t \geq t_0 \in \mathbb{R}, \ \frac{\mathsf{d}}{\mathsf{d}t} \mathsf{v}(t, \mathbf{x}_t(\cdot)) \leq -\alpha_3\left(\|\boldsymbol{x}(t)\|_2\right) \tag{A.8}
$$

*where* $\|\boldsymbol{\phi}(\cdot)\|_{\infty}^2 := \max_{-r_2 \leq \tau \leq 0} \|\boldsymbol{\phi}(\tau)\|_2^2$, *and* $\mathbf{x}_t(\cdot)$, $\boldsymbol{x}(\cdot)$ *in* (A.8) *satisfy* $\dot{\boldsymbol{x}}(t) = \boldsymbol{f}(t, \mathbf{x}_t(\cdot))$ *in* (A.5) *for almost all* $t \geq t_0$. *Moreover,* $\mathcal{K}_{\infty}$ *follows the standard definition in Khalil (2002). Note that the notation* $\widetilde{\forall}$ *means for almost all with respect to the Lebesgue measure.*

*Proof.* The proof here is based on the procedure in Theorem 2.1 of Section 5.1 in Hale & Lunel (1993), and Theorem 1.3 in Gu *et al.* (2003). To prove the uniform stability of the trivial solution, let

$$
\mathbb{R}_{\geq 0} \ni \epsilon \mapsto \delta(\epsilon) = 1/2 \min\left(\epsilon, \alpha_2^{-1}\left(\alpha_1(\epsilon)\right)\right) \tag{A.9}
$$

where $\alpha_2^{-1}(\cdot)$ is well defined since $\alpha_2(\cdot) \in \mathcal{K}_{\infty}$. It is obvious that $\delta(\cdot) \in \mathcal{K}_{\infty}$ and satisfies $\forall \epsilon > 0, 0 < \delta(\epsilon) < \epsilon$ and $\delta(\epsilon) < \alpha_2^{-1}\left(\alpha_1(\epsilon)\right)$ which further implies that

$$
\forall \epsilon > 0, \quad \alpha_2\left(\delta(\epsilon)\right) < \alpha_1(\epsilon) \tag{A.10}
$$

since $\alpha_2(\cdot) \in \mathcal{K}_\infty$. By (A.8), it is true that $\widetilde{\forall} t \geq t_0 \in \mathbb{R}$, $\dot{\mathsf{v}}(t, \mathbf{x}_t(\cdot)) \leq 0$. Now applying the fundamental theorem of calculus for the Lebesgue integrals to the previous proposition, we have

$$\begin{aligned} \forall t_0 \in \mathbb{R}, \forall t \geq t_0, \forall \boldsymbol{\phi}(\cdot) \in \mathbb{C}([-r,0] \fatsemi \mathbb{R}^n),\ & \int_{t_0}^{t} \dot{\mathsf{v}}(t, \mathbf{x}_t(\cdot)) \mathsf{d}\tau \\ & = \mathsf{v}\left(t, \mathbf{x}_t(\cdot)\right) - \mathsf{v}\big(t_0, \mathbf{x}_{t_0}(\cdot)\big) = \mathsf{v}\left(t, \mathbf{x}_t(\cdot)\right) - \mathsf{v}\big(t_0, \boldsymbol{\phi}(\cdot)\big) \leq 0 \end{aligned} \tag{A.11}$$

which further implies that $\forall t_0 \in \mathbb{R}$, $\forall t \geq t_0, \forall \epsilon > 0$, $\forall \boldsymbol{\phi}(\cdot) \in \mathbb{C}_{\delta(\epsilon)}([-r,0] \fatsemi \mathbb{R}^n)$:

$$\alpha_1(\|\boldsymbol{x}(t)\|_2) \leq \mathsf{v}(t, \mathbf{x}_t(\cdot)) \leq \mathsf{v}(t_0, \boldsymbol{\phi}(\cdot)) \leq \alpha_2\left(\|\boldsymbol{\phi}(\cdot)\|_\infty\right) < \alpha_2(\delta(\epsilon)) < \alpha_1(\epsilon) \tag{A.12}$$

by (A.7) and (A.10), where $\mathbb{C}_{\delta(\epsilon)}([-r,0] \fatsemi \mathbb{R}^n) := \left\{\boldsymbol{\phi}(\cdot) \in \mathbb{C}([-r,0] \fatsemi \mathbb{R}^n) : \|\boldsymbol{\phi}(\cdot)\|_\infty < \delta(\epsilon)\right\}$. Therefore,

$$\forall \epsilon > 0,\ \forall \boldsymbol{\phi}(\cdot) \in \mathbb{C}_{\delta(\epsilon)}([-r,0] \fatsemi \mathbb{R}^n),\ \forall t_0 \in \mathbb{R},\ \forall t \geq t_0,\ \|\boldsymbol{x}(t)\|_2 < \epsilon \tag{A.13}$$

where $\delta(\epsilon) = 1/2 \min\left(\epsilon, \alpha_2^{-1}\left(\alpha_1(\epsilon)\right)\right)$ is independent of $t_0 \in \mathbb{R}$ and $\lim_{\epsilon \to +\infty} \delta(\epsilon) = +\infty$ since $\delta(\cdot) \in \mathcal{K}_\infty$. Now (A.13) further infers that

$$\forall \epsilon > 0,\ \exists \delta > 0,\ \forall \boldsymbol{\phi}(\cdot) \in \mathbb{C}_\delta([-r,0] \fatsemi \mathbb{R}^n),\ \forall t_0 \in \mathbb{R},\ \forall t \geq t_0,\ \|\mathbf{x}_t(\cdot)\|_\infty \leq \max_{\tau \geq t_0} \|\boldsymbol{x}(\tau)\| < \epsilon \tag{A.14}$$

which shows uniform stability.

For the proof of global uniform asymptotic stability, we seek to prove it by using proof by contradiction. Note that the origin is globally uniform asymptotic stable if it is uniform stable as we have proved above and

$$\forall \eta > 0, \forall \delta > 0,\ \exists \theta \geq 0,\ \forall \boldsymbol{\phi}(\cdot) \in \mathbb{C}_\delta([-r,0] \fatsemi \mathbb{R}^n),\ \forall t_0 \in \mathbb{R},\ \forall t \geq t_0 + \theta,\ \|\mathbf{x}_t(\cdot)\|_\infty < \eta. \tag{A.15}$$

Assume that

$$\exists \epsilon > 0,\ \exists \delta > 0,\ \exists \boldsymbol{\phi}(\cdot) \in \mathbb{C}_\delta([-r,0] \fatsemi \mathbb{R}^n),\ \exists t_0 \in \mathbb{R},\ \forall t \geq t_0,\ \|\mathbf{x}_t(\cdot)\|_\infty \geq \epsilon. \tag{A.16}$$

Considering the definition $\|\mathbf{x}_t(\cdot)\|_\infty = \max_{\tau \in [-r,0]} \|\boldsymbol{x}(t+\tau)\|_2$ with (A.16), it implies

$$\exists \epsilon > 0,\ \exists \delta > 0,\ \exists \boldsymbol{\phi}(\cdot) \in \mathbb{C}_\delta([-r,0] \fatsemi \mathbb{R}^n),\ \exists t_0 \in \mathbb{R},\ \forall t \geq t_0,\ \exists \lambda \in [t-r, t],\ \|\boldsymbol{x}(\lambda)\|_2 \geq \epsilon. \tag{A.17}$$

Let $\epsilon > 0$, $\delta > 0$, $\boldsymbol{\phi}(\cdot) \in \mathbb{C}_\delta([-r,0] \fatsemi \mathbb{R}^n)$ and $t_0 \in \mathbb{R}$ in (A.17) be given, then there exists a sequence $\mathbb{N} \ni k \to t_k \in \mathbb{R}_{\geq t_0}$ such that

$$\forall k \in \mathbb{N},\ (2k-1)r \leq t_k - t_0 \leq 2kr \quad \& \quad \|\boldsymbol{x}(t_k)\|_2 \geq \epsilon. \tag{A.18}$$

On the other hand,

$$\begin{aligned} \|\boldsymbol{x}(t)\|_2 &= \left\|\boldsymbol{x}(t_k) + \int_{t_k}^{t} \dot{\boldsymbol{x}}(\tau)\mathsf{d}\tau\right\|_2 \geq \|\boldsymbol{x}(t_k)\|_2 - \left\|\int_{t_k}^{t} \dot{\boldsymbol{x}}(\tau)\mathsf{d}\tau\right\|_2 = \|\boldsymbol{x}(t_k)\|_2 - \left\|\int_{t_k}^{t} \boldsymbol{f}(\tau, \mathbf{x}_\tau(\cdot))\mathsf{d}\tau\right\|_2 \\ &\geq \|\boldsymbol{x}(t_k)\|_2 - \left\|\int_{t_k}^{t} \boldsymbol{f}(\tau, \mathbf{x}_\tau(\cdot))\mathsf{d}\tau\right\|_1 = \|\boldsymbol{x}(t_k)\|_2 - \sum_{i=1}^{n}\left|\int_{t_k}^{t} f_i(\tau, \mathbf{x}_\tau(\cdot))\mathsf{d}\tau\right| \geq \|\boldsymbol{x}(t_k)\|_2 - \left|\sum_{i=1}^{n}\int_{t_k}^{t} |f_i(\tau, \mathbf{x}_\tau(\cdot))|\,\mathsf{d}\tau\right| \\ &= \|\boldsymbol{x}(t_k)\|_2 - \left|\int_{t_k}^{t}\sum_{i=1}^{n} |f_i(\tau, \mathbf{x}_\tau(\cdot))|\,\mathsf{d}\tau\right| \geq \|\boldsymbol{x}(t_k)\|_2 - \left|\int_{t_k}^{t} \|\boldsymbol{f}(\tau, \mathbf{x}_\tau(\cdot))\|_1\,\mathsf{d}\tau\right| \end{aligned} \tag{A.19}$$

is true for all $t \geq t_0$ and $k \in \mathbb{N}$ based on the properties of Lebesgue integrals and norms. Since $\forall t \geq t_0, \forall k \in \mathbb{N}$, $\left|\int_{t_k}^{t} \|\boldsymbol{f}(\tau, \mathbf{x}_\tau(\cdot))\|_1\,\mathsf{d}\tau\right| < \left|\int_{t_k}^{t} c(\delta)\,\mathsf{d}\tau\right| = c(\delta)|t - t_k|$ by (A.6) with a given $\delta > 0$ and $\boldsymbol{\phi}(\cdot) \in \mathbb{C}_\delta([-r,0] \fatsemi \mathbb{R}^n)$, therefore we have

$$\forall k \in \mathbb{N},\ \ \forall t \in \mathcal{I}_k := \left[t_k - \frac{\epsilon}{2c(\delta)}, t_k + \frac{\epsilon}{2c(\delta)}\right],\ \ \|\boldsymbol{x}(t)\|_2 \ge \|\boldsymbol{x}(t_k)\|_2 - \left|\int_{t_k}^{t} \|\boldsymbol{f}(\tau, \mathbf{x}_\tau(\cdot))\|_1 \,\mathsf{d}\tau\right|$$

$$> \|\boldsymbol{x}(t_k)\|_2 - \left|\int_{t_k}^{t} c(\delta)\mathsf{d}\tau\right| = \|\boldsymbol{x}(t_k\|_2 - c(\delta)|t - t_k| \ge \epsilon - c(\delta)\frac{\epsilon}{2c(\delta)} = \frac{\epsilon}{2}. \tag{A.20}$$

Consequently,

$$\widetilde{\forall} t \in \mathbb{R}_{\ge t_0} \cap \bigcup_{k\in\mathbb{N}} \mathcal{I}_k,\ \frac{\mathsf{d}}{\mathsf{d}t}\mathsf{v}(\mathbf{x}_t(\cdot)) \le -\alpha_3\left(\epsilon/2\right).\ \ \&\ \ \widetilde{\forall} t \in \mathbb{R}_{\ge t_0},\ \ \frac{\mathsf{d}}{\mathsf{d}t}\mathsf{v}(\mathbf{x}_t(\cdot)) \le 0. \tag{A.21}$$

Since $c(\delta) > 0$ in $\mathcal{I}_k = [t_k - \epsilon/2c(\delta), t_k + \epsilon/2c(\delta)]$ can be made arbitrarily large for any $\delta > 0$, thus we can assume that $\bigcap_{k\in\mathbb{N}}[t_k - \epsilon/2c(\delta), t_k + \epsilon/2c(\delta)] = \varnothing$ and $t_1 - \epsilon/2c(\delta) \ge t_0$. As a result, we have

$$\forall k \in \mathbb{N},\ \mathsf{v}(t_k, \mathbf{x}_{t_k}(\cdot)) - \mathsf{v}(t_0, \boldsymbol{\phi}(\cdot)) = \int_{t_0}^{t_k} \frac{\mathsf{d}}{\mathsf{d}\tau}\mathsf{v}(\mathbf{x}_\tau(\cdot))\mathsf{d}\tau = \int_{\bigcup_{i=1}^{k-1}\mathcal{I}_i} \frac{\mathsf{d}}{\mathsf{d}\tau}\mathsf{v}(\mathbf{x}_\tau(\cdot))\mathsf{d}\tau + \int_{[t_k, t_0]\setminus\bigcup_{i=1}^{k-1}\mathcal{I}_i} \frac{\mathsf{d}}{\mathsf{d}\tau}\mathsf{v}(\mathbf{x}_\tau(\cdot))\mathsf{d}\tau$$

$$\le -\int_{\bigcup_{i=1}^{k-1}\mathcal{I}_i} \alpha_3\left(\epsilon/2\right)\mathsf{d}\tau + \int_{[t_k, t_0]\setminus\bigcup_{i=1}^{k-1}\mathcal{I}_i} 0\mathsf{d}\tau = -\sum_{i=1}^{k-1}\int_{t_i - \epsilon/2c(\delta)}^{t_i + \epsilon/2c(\delta)} \alpha_3\left(\epsilon/2\right)\mathsf{d}\tau = -\alpha_3\left(\epsilon/2\right)\frac{\epsilon}{c(\delta)}(k-1) \tag{A.22}$$

by (A.21). This further infers that

$$\forall k \in \mathbb{N},\ \mathsf{v}(t_k, \mathbf{x}_{t_k}(\cdot)) \le \mathsf{v}(t_0, \boldsymbol{\phi}(\cdot)) - \alpha_3\left(\epsilon/2\right)\frac{\epsilon}{c(\delta)}(k-1) \le \alpha_2\left(\|\boldsymbol{\phi}(\cdot)\|_\infty\right) - \alpha_3\left(\epsilon/2\right)\frac{\epsilon}{c(\delta)}(k-1)$$

$$< \alpha_2(\delta) - \alpha_3\left(\epsilon/2\right)\frac{\epsilon}{c(\delta)}(k-1) \tag{A.23}$$

by (A.7) and the fact that $\|\boldsymbol{\phi}(\cdot)\|_\infty < \delta$ and $\alpha_2(\cdot) \in \mathcal{K}_\infty$. Note that

$$\alpha_2(\delta) - \alpha_3\left(\epsilon/2\right)\frac{\epsilon}{c(\delta)}(k-1) < 0 \iff \frac{\alpha_2(\delta)}{\alpha_3\left(\epsilon/2\right)}\frac{c(\delta)}{\epsilon} + 1 < k. \tag{A.24}$$

Let $\kappa(\epsilon, \delta) = \left\lceil \frac{\alpha_2(\delta)}{\alpha_3\left(\epsilon/2\right)}\frac{c(\delta)}{\epsilon} \right\rceil + 1$. Hence we have $\forall k > \kappa(\epsilon, \delta)$, $\mathsf{v}(t_k, \mathbf{x}_{t_k}(\cdot)) < 0$ by (A.23) which is a contradiction considering (29). As a result, (A.16) cannot be true for $t_k$ with any $k > \kappa(\epsilon, \delta)$, which implies that $\exists k \le \kappa(\epsilon, \delta)$, $\|\mathbf{x}_{t_k}(\cdot)\|_\infty < \epsilon$. This further infers that

$$\forall \epsilon > 0,\ \forall \delta > 0,\ \forall \boldsymbol{\phi}(\cdot) \in \mathbb{C}_\delta([-r, 0] ; \mathbb{R}^n),\ \forall t_0 \in \mathbb{R},\ \exists \theta \in [t_0, t_0 + 2r\kappa(\epsilon, \delta)],\ \|\mathbf{x}_\theta(\cdot)\|_\infty < \epsilon \tag{A.25}$$

considering (A.18).

Let $\epsilon > 0$ in (A.25) to be

$$\epsilon(\eta) = 1/3 \min\left(\eta, \alpha_2^{-1}\left(\alpha_1(\eta)\right)\right) \tag{A.26}$$

with a given $\eta > 0$, and assume $\boldsymbol{\phi}(\cdot)$, $t_0$, $\theta$ in (A.25) are also given. Note that (A.26) guarantees $\epsilon(\cdot) \in \mathcal{K}_\infty$ and $\alpha_2(\epsilon(\eta)) < \alpha_1(\eta)$ for any $\eta > 0$ similar to the property in (A.10).

Now let $\boldsymbol{\psi}\big(t, t_0, \boldsymbol{\phi}(\cdot)\big)(\cdot) \in \mathbb{C}\left([-r, 0] ; \mathbb{R}^n\right)$ denotes the unique solution of (A.5) with explicit dependence of $t_0$ and $\boldsymbol{\phi}(\cdot)$. Note that $\forall t \ge t_0$, $\boldsymbol{\psi}\big(t, t_0, \boldsymbol{\phi}(\cdot)\big)(\cdot) = \mathbf{x}_t(\cdot)$. By using the cocyclic property[8] of $\boldsymbol{\psi}(t, t_0, \boldsymbol{\phi}(\cdot))(\cdot)$, we have $\forall \eta > 0$, $\forall \delta > 0$, $\forall \boldsymbol{\phi}(\cdot) \in \mathbb{C}_\delta([-r, 0] ; \mathbb{R}^n)$, $\forall t_0 \in \mathbb{R}$, $\forall t \in [\theta, +\infty) \supseteq [t_0 + 2r\kappa(\epsilon(\eta), \delta)), +\infty)$

$$\boldsymbol{\psi}\left(t, \theta, \mathbf{x}_\theta(\cdot)\right)(\cdot) = \boldsymbol{\psi}\big(t, t_0, \boldsymbol{\phi}(\cdot)\big)(\cdot) = \mathbf{x}_t(\cdot). \tag{A.27}$$

[8] For the cocyclic property of dynamical systems, see eq.(6) in Chapter 2 of Hinrichsen & Pritchard (2005)

By (A.27) and (A.7), we have $\forall \eta > 0$, $\forall \delta > 0$, $\forall t_0 \in \mathbb{R}$, $\forall \boldsymbol{\phi}(\cdot) \in \mathbb{C}_\delta([-r, 0] ; \mathbb{R}^n)$, $\forall t \in [\theta, +\infty) \supseteq [t_0 + 2r\kappa(\epsilon(\eta), \delta), +\infty)$

$$\alpha_1(\|\boldsymbol{x}(t)\|_2) \leq \mathsf{v}(\mathbf{x}_t(\cdot)) = \mathsf{v}\left(\boldsymbol{\psi}\Big[t, \theta, \mathbf{x}_\theta(\cdot)\Big](\cdot)\right) \leq \mathsf{v}\left(\mathbf{x}_\theta(\cdot)\right) \leq \alpha_2(\epsilon(\eta)) < \alpha_1(\eta) \tag{A.28}$$

which further implies that $\|\boldsymbol{x}(t)\|_2 < \eta$ since $\alpha_1(\cdot) \in \mathcal{K}_\infty$. Because $2r\kappa(\epsilon(\eta), \delta)$ is independent of $t_0$, hence one can coonclude that

$$\forall \eta > 0,\ \forall \delta > 0,\ \exists \tau = 2r\kappa(\epsilon(\eta), \delta) > 0,\ \forall \boldsymbol{\phi}(\cdot) \in \mathbb{C}_\delta([-r, 0] ; \mathbb{R}^n),\ \forall t_0 \in \mathbb{R},\ \forall t \geq t_0 + \tau, \|\boldsymbol{x}(t)\|_2 < \eta. \tag{A.29}$$

This shows the global uniform asymptotic stability in defined in (A.12) since the uniform stability has been proved with the $\delta(\cdot) \in \mathcal{K}_\infty$ in (A.9) satisfying $\lim_{\epsilon \to +\infty} \delta(\epsilon) = +\infty$. This finishes the proof of this theorem. ■

## Appendix B. Two integral inequalities

**Lemma 5.** *Given $\varpi(\cdot) \in \mathbb{M}_{\boldsymbol{\mathcal{L}}(\mathcal{K})/\boldsymbol{\mathcal{B}}(\mathbb{R})}(\mathcal{K} ; \mathbb{R}_{\geq 0})$ and assume $\varpi(\cdot)$ has only countably infinite or finite number of zero values, where $\mathcal{K} \in \boldsymbol{\mathcal{L}}(\mathbb{R})$ and its Lebesgue measure is non-zero. Suppose $U \in \mathbb{S}^n_{\succeq 0}$ and $\mathsf{f}(\cdot) \in \mathbb{L}^2_\varpi\left(\mathcal{K} ; \mathbb{R}^d\right)$ satisfying*

$$\int_{\mathcal{K}} \varpi(\tau)\mathsf{f}(\tau)\mathsf{f}^\top(\tau)\mathrm{d}\tau \succ 0, \tag{B.1}$$

*then we have*

$$\int_{\mathcal{K}} \varpi(\tau)\boldsymbol{x}^\top(\tau)U\boldsymbol{x}(\tau)\mathrm{d}\tau \geq \int_{\mathcal{K}} \varpi(\tau)\boldsymbol{x}^\top(\tau)F^\top(\tau)\mathrm{d}\tau \left(\mathsf{F}^{-1} \otimes U\right)\int_{\mathcal{K}} \varpi(\tau)F(\tau)\boldsymbol{x}(\tau)\mathrm{d}\tau \tag{B.2}$$

*for all $\boldsymbol{x}(\cdot) \in \mathbb{L}^2_\varpi(\mathcal{K} ; \mathbb{R}^n)$, where $n; d \in \mathbb{N}$ and $F(\tau) = \mathsf{f}(\tau) \otimes I_n$ and $\mathsf{F} = \int_{\mathcal{K}} \varpi(\tau)\mathsf{f}(\tau)\mathsf{f}^\top(\tau)\mathrm{d}\tau$ and*

$$\mathbb{L}^2_\varpi\left(\mathcal{K} ; \mathbb{R}^d\right) = \left\{\boldsymbol{\phi}(\cdot) \in \mathbb{M}_{\boldsymbol{\mathcal{L}}(\mathcal{K})/\boldsymbol{\mathcal{B}}(\mathbb{R}^d)}\left(\mathcal{K} ; \mathbb{R}^d\right) : \|\boldsymbol{\phi}(\cdot)\|_\varpi < \infty\right\} \tag{B.3}$$

*with $\|\boldsymbol{\phi}(\cdot)\|^2_\varpi := \int_{\mathcal{K}} \varpi(\tau)\boldsymbol{\phi}^\top(\tau)\boldsymbol{\phi}(\tau)\mathrm{d}\tau$.*

*Proof.* See Theorem 1 in Feng & Nguang (2018). Note that F in (B.2) is defined differently compared to the definition of F in Feng & Nguang (2018). ■

**Lemma 6.** *Given $\mathcal{K} = [a, b]$ with $0 \leq a < b$ and . Assume $U \in \mathbb{S}^n_{\succeq 0}$ with $n \in \mathbb{N}$ and $\mathsf{f}(\tau) := \mathbf{Col}_{i=1}^d f_i(\tau) \in \mathbb{L}^2_\varpi\left([a, b] ; \mathbb{R}^d\right)$ satisfying*

$$\int_a^b \varpi(\tau)\mathsf{f}(\tau)\mathsf{f}^\top(\tau)\mathrm{d}\tau \succ 0, \tag{B.4}$$

*then we have*

$$\begin{aligned}
\int_a^b \varpi(\tau)\boldsymbol{x}^\top(\tau)U\boldsymbol{x}(\tau)\mathrm{d}\tau &\geq [*]\left(\begin{bmatrix} U & Y \\ * & U \end{bmatrix} \otimes \mathsf{F}^{-1}\right)\begin{bmatrix} \int_\varrho^b \left(I_n \otimes \mathsf{f}(\tau)\right)\boldsymbol{x}(\tau)\varpi(\tau)\mathrm{d}\tau \\ \int_a^\varrho \left(I_n \otimes \mathsf{f}(\tau)\right)\boldsymbol{x}(\tau)\varpi(\tau)\mathrm{d}\tau \end{bmatrix} \\
&= [*]\left(\begin{bmatrix} \mathsf{K}_{(d,n)} & \mathsf{O}_{dn} \\ * & \mathsf{K}_{(d,n)} \end{bmatrix}\left(\begin{bmatrix} U & Y \\ * & U \end{bmatrix} \otimes \mathsf{F}^{-1}\right)\begin{bmatrix} \mathsf{K}_{(n,d)} & \mathsf{O}_{dn} \\ * & \mathsf{K}_{(n,d)} \end{bmatrix}\right)\begin{bmatrix} \int_\varrho^b \left(\mathsf{f}(\tau) \otimes I_n\right)\boldsymbol{x}(\tau)\varpi(\tau)\mathrm{d}\tau \\ \int_a^\varrho \left(\mathsf{f}(\tau) \otimes I_n\right)\boldsymbol{x}(\tau)\varpi(\tau)\mathrm{d}\tau \end{bmatrix}
\end{aligned} \tag{B.5}$$

*for all $\boldsymbol{x}(\cdot) \in \mathbb{L}^2_\varpi(\mathcal{K} ; \mathbb{R}^n)$, $\varrho \in [a, b]$ and for any $Y \in \mathbb{R}^{n \times n}$ satisfying $\left[\begin{smallmatrix} U & Y \\ * & U \end{smallmatrix}\right] \succeq 0$, where $\mathsf{F} = \int_a^b \varpi(\tau)\mathsf{f}(\tau)\mathsf{f}^\top(\tau)\mathrm{d}\tau$.*

*Proof.* The proof is based on the insights illustrated in Section 4.1 of Seuret *et al.* (2016). Consider the equality

$$\begin{aligned}\int_a^b \varpi(\tau)\boldsymbol{x}^\top(\tau)U\boldsymbol{x}(\tau)\mathsf{d}\tau &= \int_\varrho^b \varpi(\tau)\begin{bmatrix}\boldsymbol{x}(\tau)\\ \mathbf{0}_n\end{bmatrix}^\top \begin{bmatrix}U & Y\\ * & U\end{bmatrix}\begin{bmatrix}\boldsymbol{x}(\tau)\\ \mathbf{0}_n\end{bmatrix}\mathsf{d}\tau \\ &+ \int_a^\varrho \varpi(\tau)\begin{bmatrix}\mathbf{0}_n\\ \boldsymbol{x}(\tau)\end{bmatrix}^\top \begin{bmatrix}U & Y\\ * & U\end{bmatrix}\begin{bmatrix}\mathbf{0}_n\\ \boldsymbol{x}(\tau)\end{bmatrix}\mathsf{d}\tau = \int_a^b \boldsymbol{y}^\top(\tau)\begin{bmatrix}U & Y\\ * & U\end{bmatrix}\boldsymbol{y}(\tau)\mathsf{d}\tau \end{aligned} \tag{B.6}$$

which holds for any $Y \in \mathbb{R}^{n\times n}$ with

$$\mathbb{R}^{2n} \ni \boldsymbol{y}(\tau) := \begin{cases}\begin{bmatrix}\boldsymbol{x}(\tau)\\ \mathbf{0}_n\end{bmatrix}, & \forall\tau\in[\varrho,b] \\ \begin{bmatrix}\mathbf{0}_n\\ \boldsymbol{x}(\tau)\end{bmatrix}, & \forall\tau\in[a,\varrho],\end{cases} \qquad \varrho\in[a,b]. \tag{B.7}$$

Let $Y \in \mathbb{R}^{n\times n}$ satisfying $\left[\begin{smallmatrix}U & Y\\ * & U\end{smallmatrix}\right] \succeq 0$, then one can apply (B.2) with (A.3)–(A.4) to the rightmost integral in (B.6) with $\mathcal{K}=[a,b]$ and $\mathsf{f}(\cdot)\in\mathbb{L}^2_\varpi\left(\mathcal{K} \mathbin{;} \mathbb{R}^d\right)$ satisfying (B.4). Then we have

$$\begin{aligned}\int_a^b \varpi(\tau)\boldsymbol{x}^\top(\tau)U\boldsymbol{x}(\tau)\mathsf{d}\tau &= \int_a^b \varpi(\tau)\boldsymbol{y}^\top(\tau)\begin{bmatrix}U & Y\\ * & U\end{bmatrix}\boldsymbol{y}(\tau)\mathsf{d}\tau \\ &\geq [*]\left(\mathsf{F}^{-1}\otimes\begin{bmatrix}U & Y\\ * & U\end{bmatrix}\right)\left(\int_a^b \varpi(\tau)\left(\mathsf{f}(\tau)\otimes I_{2n}\right)\boldsymbol{y}(\tau)\mathsf{d}\tau\right) \\ &= [*]\left(\mathsf{F}^{-1}\otimes\begin{bmatrix}U & Y\\ * & U\end{bmatrix}\right)\left(\int_a^b \varpi(\tau)\mathsf{K}_{2n,d}\left(I_{2n}\otimes\mathsf{f}(\tau)\right)\boldsymbol{y}(\tau)\mathsf{d}\tau\right) \\ &= \int_a^b \varpi(\tau)\boldsymbol{y}^\top(\tau)\left(I_{2n}\otimes\mathsf{f}^\top(\tau)\right)\mathsf{d}\tau\left(\begin{bmatrix}U & Y\\ * & U\end{bmatrix}\otimes\mathsf{F}^{-1}\right)\int_a^b \varpi(\tau)\left(I_{2n}\otimes\mathsf{f}(\tau)\right)\boldsymbol{y}(\tau)\mathsf{d}\tau\end{aligned} \tag{B.8}$$

where $\mathsf{F}=\int_a^b \varpi(\tau)\mathsf{f}(\tau)\mathsf{f}^\top(\tau)\mathsf{d}\tau$. Furthermore, it follows that

$$\begin{aligned}\int_a^b \left(I_{2n}\otimes\mathsf{f}(\tau)\right)\boldsymbol{y}(\tau)\varpi(\tau)\mathsf{d}\tau &= \int_\varrho^b \begin{bmatrix}I_n\otimes\mathsf{f}(\tau) & \mathsf{O}_{dn}\\ \mathsf{O}_{dn} & I_n\otimes\mathsf{f}(\tau)\end{bmatrix}\begin{bmatrix}\boldsymbol{x}(\tau)\\ \mathbf{0}_n\end{bmatrix}\varpi(\tau)\mathsf{d}\tau \\ &+ \int_a^\varrho \begin{bmatrix}I_n\otimes\mathsf{f}(\tau) & \mathsf{O}_{dn}\\ \mathsf{O}_{dn} & I_n\otimes\mathsf{f}(\tau)\end{bmatrix}\begin{bmatrix}\mathbf{0}_n\\ \boldsymbol{x}(\tau)\end{bmatrix}\varpi(\tau)\mathsf{d}\tau = \begin{bmatrix}\int_\varrho^b \left[I_n\otimes\mathsf{f}(\tau)\right]\boldsymbol{x}(\tau)\varpi(\tau)\mathsf{d}\tau\\ \int_a^\varrho \left[I_n\otimes\mathsf{f}(\tau)\right]\boldsymbol{x}(\tau)\varpi(\tau)\mathsf{d}\tau\end{bmatrix}\end{aligned} \tag{B.9}$$

by the definition of the Kronecker product. Substituting (B.9) into (B.8) and using (A.4) yield (B.5). ■

> **Remark 15.** Note that the value of $\mathsf{F}$ in (B.5) is related to the values of $a$ and $b$, and not related to the value of $\varrho$. Thus it means that $\varrho$ can be a function of any kind as long as its value is bounded by $[a,b]$. This property enables us to deal with time-varying delays, and derive tractable dissipative conditions in the next section.

### Appendix C. Proof of Theorem 1

The proof of Theorem 1 is via the construction of

$$\begin{aligned}\mathsf{v}(\mathbf{x}_t(\cdot)) &= \boldsymbol{\eta}^\top(t)\begin{bmatrix}P_1 & P_2\\ * & P_3\end{bmatrix}\boldsymbol{\eta}(t) + \int_{-r_1}^0 \boldsymbol{x}^\top(t+\tau)\left[Q_1+(\tau+r_1)R_1\right]\boldsymbol{x}(t+\tau)\mathsf{d}\tau \\ &+ \int_{-r_2}^{-r_1} \boldsymbol{x}^\top(t+\tau)\left[Q_2+(\tau+r_2)R_2\right]\boldsymbol{x}(t+\tau)\mathsf{d}\tau\end{aligned} \tag{C.1}$$

where $\mathbf{x}_t(\cdot)$ follows the same definition in (31), and $P_1 \in \mathbb{S}^n$, $P_2 \in \mathbb{R}^{n\times\varrho}$, $P_3 \in \mathbb{S}^{\varrho}$ with $\varrho = (d_1 + d_2)n$, and $Q_1; Q_2; R_1; R_2 \in \mathbb{S}^n$ and

$$\boldsymbol{\eta}(t) := \mathsf{\textbf{Col}}\left[\boldsymbol{x}(t),\ \int_{-r_1}^{0}\left(\sqrt{\mathsf{F}_1^{-1}}\boldsymbol{f}_1(\tau)\otimes I_n\right)\boldsymbol{x}(t+\tau)\mathsf{d}\tau,\ \int_{-r_2}^{-r_1}\left(\sqrt{\mathsf{F}_2^{-1}}\boldsymbol{f}_2(\tau)\otimes I_n\right)\boldsymbol{x}(t+\tau)\mathsf{d}\tau\right] \tag{C.2}$$

with $\mathsf{F}_1 = \int_{-r_1}^{0}\boldsymbol{f}_1(\tau)\boldsymbol{f}_1^{\top}(\tau)\mathsf{d}\tau$ and $\mathsf{F}_2 = \int_{-r_2}^{-r_1}\boldsymbol{f}_2(\tau)\boldsymbol{f}_2^{\top}(\tau)\mathsf{d}\tau$. Note that given the conditions in (8)–(9), both $\sqrt{\mathsf{F}_1^{-1}}$ and $\sqrt{\mathsf{F}_2^{-1}}$ are well defined.

We will first prove this theorem for the case of $r_2 > r_1 > 0$. Then the synthesis conditions for the cases of $r_1 = r_2 > 0$ and $r_1 = 0$; $r_2 > 0$ can be easily obtained based on the synthesis condition for $r_2 > r_1 > 0$, respectively.

Now given $t_0 \in \mathbb{R}$ in (19) with $r_2 > r_1 > 0$, differentiating $\mathsf{v}(\mathbf{x}_t(\cdot))$ along the trajectory of (19) and consider (32) produces

$$\begin{aligned}
&\widetilde{\forall} t \geq t_0,\ \ \dot{\mathsf{v}}(\mathbf{x}_t(\cdot)) - \mathsf{s}(\boldsymbol{z}(t), \boldsymbol{w}(t)) \\
&= \boldsymbol{\chi}^{\top}(t)\,\mathsf{\textbf{Sy}}\left(\begin{bmatrix} \mathsf{O}_{2n,n} & \mathsf{O}_{2n,\varrho} \\ I_n & \mathsf{O}_{n,\varrho} \\ \mathsf{O}_{\kappa n,n} & \widehat{I}^{\top} \\ \mathsf{O}_{q,n} & \mathsf{O}_{q,\varrho} \end{bmatrix}\begin{bmatrix} P_1 & P_2 \\ * & P_3 \end{bmatrix}\begin{bmatrix} \mathbf{A} + \mathbf{B}_1\left[\left(I_{\widehat{3}+\kappa}\otimes K\right)\oplus \mathsf{O}_q\right] \\ \left[\widehat{\mathbf{F}}\otimes I_n \quad \mathsf{O}_{\varrho,q}\right] \end{bmatrix} - \begin{bmatrix} \mathsf{O}_{(3n+\kappa n),m} \\ J_2^{\top} \end{bmatrix}\boldsymbol{\Sigma}\right)\boldsymbol{\chi}(t) \\
&+ \boldsymbol{x}^{\top}(t)\left(Q_1 + r_1R_1\right)\boldsymbol{x}(t) - \boldsymbol{x}^{\top}(t-r_2)Q_2\boldsymbol{x}(t-r_2) - \boldsymbol{x}^{\top}(t-r_1)\left(Q_1 - Q_2 - r_3R_2\right)\boldsymbol{x}(t-r_1) \\
&- \boldsymbol{w}^{\top}(t)J_3\boldsymbol{w}(t) - \int_{-r_1}^{0}\boldsymbol{x}^{\top}(t+\tau)R_1\boldsymbol{x}(t+\tau)\mathsf{d}\tau - \int_{-r_2}^{-r_1}\boldsymbol{x}^{\top}(t+\tau)R_2\boldsymbol{x}(t+\tau)\mathsf{d}\tau \\
&- \boldsymbol{\chi}^{\top}(t)\boldsymbol{\Sigma}^{\top}\widetilde{J}^{\top}J_1^{-1}\widetilde{J}\boldsymbol{\Sigma}\boldsymbol{\chi}(t)
\end{aligned} \tag{C.3}$$

where $\boldsymbol{\chi}(t)$ is given in (24) and $\boldsymbol{\Sigma}$, $\widehat{I}$ and $\widehat{\mathbf{F}}$ are defined in the statements of Theorem 1. Note that the expression of $\widehat{\mathbf{F}}$ in (39) is obtained by the relations

$$\begin{aligned}
&\int_{-r_1}^{0}\left(\sqrt{\mathsf{F}_1^{-1}}\boldsymbol{f}_1(\tau)\otimes I_n\right)\dot{\boldsymbol{x}}(t+\tau)\mathsf{d}\tau = \left(\sqrt{\mathsf{F}_1^{-1}}\boldsymbol{f}_1(0)\otimes I_n\right)\boldsymbol{x}(t) \\
&- \left(\sqrt{\mathsf{F}_1^{-1}}\boldsymbol{f}_1(-r_1)\otimes I_n\right)\boldsymbol{x}(t-r_1) - \left(\sqrt{\mathsf{F}_1^{-1}}M_1\sqrt{\mathsf{G}_1}\otimes I_n\right)\int_{-r_1}^{0}\left(\sqrt{\mathsf{G}_1^{-1}}\widehat{\boldsymbol{f}}_1(\tau)\otimes I_n\right)\boldsymbol{x}(t+\tau)\mathsf{d}\tau
\end{aligned} \tag{C.4}$$

$$\begin{aligned}
\int_{-r_2}^{-r_1}\left(\sqrt{\mathsf{F}_2^{-1}}\boldsymbol{f}_2(\tau)\otimes I_n\right)\dot{\boldsymbol{x}}(t+\tau)\mathsf{d}\tau &= \left(\sqrt{\mathsf{F}_2^{-1}}\boldsymbol{f}_2(-r_1)\otimes I_n\right)\boldsymbol{x}(t-r_1) - \left(\sqrt{\mathsf{F}_2^{-1}}\boldsymbol{f}_2(-r_2)\otimes I_n\right)\boldsymbol{x}(t-r_2) \\
&- \left(\sqrt{\mathsf{F}_2^{-1}}M_2\sqrt{\mathsf{G}_2}\otimes I_n\right)\int_{-r(t)}^{-r_1}\left(\sqrt{\mathsf{G}_2^{-1}}\widehat{\boldsymbol{f}}_2(\tau)\otimes I_n\right)\boldsymbol{x}(t+\tau)\mathsf{d}\tau \\
&- \left(\sqrt{\mathsf{F}_2^{-1}}M_2\sqrt{\mathsf{G}_2}\otimes I_n\right)\int_{-r_2}^{-r(t)}\left(\sqrt{\mathsf{G}_2^{-1}}\widehat{\boldsymbol{f}}_2(\tau)\otimes I_n\right)\boldsymbol{x}(t+\tau)\mathsf{d}\tau
\end{aligned} \tag{C.5}$$

which are derived via (7)–(10) and (A.1)–(A.2). On the other hand, the structure of $\widehat{I}$ in (C.3) is obtained based on the identities

$$\boldsymbol{f}_1(\tau) = \begin{bmatrix}\mathsf{O}_{d_1,\delta_1} & I_{d_1}\end{bmatrix}\widehat{\boldsymbol{f}}_1(\tau),\quad \boldsymbol{f}_2(\tau) = \begin{bmatrix}\mathsf{O}_{d_2,\delta_2} & I_{d_2}\end{bmatrix}\widehat{\boldsymbol{f}}_2(\tau) \tag{C.6}$$

$$
\begin{bmatrix} \int_{-r_1}^{0} \left( \sqrt{\mathsf{F}_1^{-1}} \boldsymbol{f}_1(\tau) \otimes I_n \right) \boldsymbol{x}(t+\tau)\mathsf{d}\tau \\ \int_{-r_2}^{-r_1} \left( \sqrt{\mathsf{F}_2^{-1}} \boldsymbol{f}_2(\tau) \otimes I_n \right) \boldsymbol{x}(t+\tau)\mathsf{d}\tau \end{bmatrix} = \widehat{I} \begin{bmatrix} \int_{-r_1}^{0} \left( \sqrt{\mathsf{G}_1^{-1}} \widehat{\boldsymbol{f}}_1(\tau) \otimes I_n \right) \boldsymbol{x}(t+\tau)\mathsf{d}\tau \\ \int_{-r(t)}^{-r_1} \left( \sqrt{\mathsf{G}_2^{-1}} \widehat{\boldsymbol{f}}_2(\tau) \otimes I_n \right) \boldsymbol{x}(t+\tau)\mathsf{d}\tau \\ \int_{-r_2}^{-r(t)} \left( \sqrt{\mathsf{G}_2^{-1}} \widehat{\boldsymbol{f}}_2(\tau) \otimes I_n \right) \boldsymbol{x}(t+\tau)\mathsf{d}\tau \end{bmatrix} \tag{C.7}
$$

in light of the form of $\boldsymbol{\eta}(t)$ in (C.2) and $\boldsymbol{\chi}(t)$ in (24) and the property of the Kronecker product in (A.2). Note that also the parameters $\mathbf{A}$, $\mathbf{B}_1$, $\mathbf{C}$ and $\mathbf{B}_2$ in (C.3) are given in (20)–(23).

Let $R_1 \succeq 0$ and $\left[\begin{smallmatrix} R_2 & Y \\ * & R_2 \end{smallmatrix}\right] \succeq 0$ with $Y \in \mathbb{R}^{n \times n}$. Now apply (B.2) and (B.5) with $\varpi(\tau) = 1$ and $\mathsf{f}(\tau) = \sqrt{\mathsf{G}_1^{-1}} \widehat{\boldsymbol{f}}_1(\tau)$, $\mathsf{f}(\tau) = \sqrt{\mathsf{G}_2^{-1}} \widehat{\boldsymbol{f}}_2(\tau)$ to the integral terms $\int_{-r_1}^{0} \boldsymbol{x}^\top(t+\tau)R_1\boldsymbol{x}(t+\tau)\mathsf{d}\tau$ and $\int_{-r_2}^{-r_1} \boldsymbol{x}^\top(t+\tau)R_2\boldsymbol{x}(t+\tau)\mathsf{d}\tau$ in (C.3), respectively. Then we have

$$
\int_{-r_1}^{0} \boldsymbol{x}^\top(t+\tau)R_1\boldsymbol{x}(t+\tau)\mathsf{d}\tau \geq [*]\left(I_{\kappa_1} \otimes R_1\right)\left[\int_{-r_1}^{0} \left( \sqrt{\mathsf{G}_1^{-1}} \widehat{\boldsymbol{f}}_1(\tau) \otimes I_n \right) \boldsymbol{x}(t+\tau)\mathsf{d}\tau\right] \tag{C.8}
$$

$$
\begin{aligned}
\int_{-r_2}^{-r_1} \boldsymbol{x}^\top(t+\tau)R_2\boldsymbol{x}(t+\tau)\mathsf{d}\tau &\geq [*]\left(\begin{bmatrix} R_2 & Y \\ * & R_2 \end{bmatrix} \otimes I_{\kappa_2}\right) \begin{bmatrix} \int_{-r(t)}^{-r_1} \left( I_n \otimes \sqrt{\mathsf{G}_2^{-1}} \widehat{\boldsymbol{f}}_2(\tau) \right) \boldsymbol{x}(t+\tau)\mathsf{d}\tau \\ \int_{-r_2}^{-r(t)} \left( I_n \otimes \sqrt{\mathsf{G}_2^{-1}} \widehat{\boldsymbol{f}}_2(\tau) \right) \boldsymbol{x}(t+\tau)\mathsf{d}\tau \end{bmatrix} \\
&= [*]\left(\begin{bmatrix} \mathsf{K}_{(\kappa_2,n)} & \mathsf{O}_{\kappa_2 n} \\ * & \mathsf{K}_{(\kappa_2,n)} \end{bmatrix}\left(\begin{bmatrix} R_2 & Y \\ * & R_2 \end{bmatrix} \otimes I_{\kappa_2}\right)\begin{bmatrix} \mathsf{K}_{(n,\kappa_2)} & \mathsf{O}_{\kappa_2 n} \\ * & \mathsf{K}_{(n,\kappa_2)} \end{bmatrix}\right) \begin{bmatrix} \int_{-r(t)}^{-r_1} \left( \sqrt{\mathsf{G}_2^{-1}} \widehat{\boldsymbol{f}}_2(\tau) \otimes I_n \right) \boldsymbol{x}(t+\tau)\mathsf{d}\tau \\ \int_{-r_2}^{-r(t)} \left( \sqrt{\mathsf{G}_2^{-1}} \widehat{\boldsymbol{f}}_2(\tau) \otimes I_n \right) \boldsymbol{x}(t+\tau)\mathsf{d}\tau \end{bmatrix}.
\end{aligned} \tag{C.9}
$$

Given the definition of $\mathbb{1}$ and $\widehat{\mathbb{1}}$ in (25) and $\widehat{\mathsf{O}}$ in (24) for the case of $r_2 > r_1 > 0$, applying (C.8)–(C.9) to (C.3) with (34) produces

$$
\widetilde{\forall} t \geq t_0,\ \ \dot{\mathsf{v}}(\mathbf{x}_t(\cdot)) - \mathsf{s}(\boldsymbol{z}(t), \boldsymbol{w}(t)) \leq \boldsymbol{\chi}^\top(t)\left(\boldsymbol{\Psi} - \boldsymbol{\Sigma}^\top \widetilde{J}^\top J_1^{-1} \widetilde{J} \boldsymbol{\Sigma}\right)\boldsymbol{\chi}(t) \tag{C.10}
$$

where $\boldsymbol{\Psi}$ is given in (36) and $\boldsymbol{\chi}(t)$ is given in (24). Now it is obvious to conclude that if (34) and $\boldsymbol{\Psi} - \boldsymbol{\Sigma}^\top \widetilde{J}^\top J_1^{-1} \widetilde{J} \boldsymbol{\Sigma} \prec 0$ are true, then

$$
\exists \epsilon_3 > 0 : \widetilde{\forall} t \geq t_0,\ \ \dot{\mathsf{v}}(\mathbf{x}_t(\cdot)) - \mathsf{s}(\boldsymbol{z}(t), \boldsymbol{w}(t)) \leq -\epsilon_3 \left\|\boldsymbol{x}(t)\right\|_2. \tag{C.11}
$$

Moreover, assuming $\boldsymbol{w}(t) \equiv \mathbf{0}_q$, one can also obtain

$$
\exists \epsilon_3 > 0,\ \widetilde{\forall} t \geq t_0,\ \dot{\mathsf{v}}(\mathbf{x}_t(\cdot)) \leq -\epsilon_3 \left\|\boldsymbol{x}(t)\right\|_2 \tag{C.12}
$$

by the structure of $\boldsymbol{\Psi}$ with the fact that $\boldsymbol{\Psi} \prec 0$ and the elements in $\boldsymbol{\chi}(t)$ considering the properties of quadratic forms. Note that $\mathbf{x}_t(\cdot)$ in (C.12) is in line with the definition of $\mathbf{x}_t(\cdot)$ in (30). As a result, there exists a functional in (C.1) satisfying (31) and (30) if (34) and $\boldsymbol{\Psi} - \boldsymbol{\Sigma}^\top \widetilde{J}^\top J_1^{-1} \widetilde{J} \boldsymbol{\Sigma} \prec 0$ are feasible for some matrices. Finally, applying the Schur complement to $\boldsymbol{\Psi} - \boldsymbol{\Sigma}^\top \widetilde{J}^\top J_1^{-1} \widetilde{J} \boldsymbol{\Sigma} \prec 0$ with (34) and $J_1^{-1} \prec 0$ gives the equivalent condition in (35). Therefore we have proved that the existence of the feasible solutions of (34) and (35) infer the existence of a functional (C.1) and $\epsilon_3 > 0$ satisfying (31) and (30).

Now we start to show that if (33) and (34) are feasible for some matrices, then there exist $\epsilon_1 > 0$ and $\epsilon_2 > 0$ such that (C.1) satisfies (29). Let $\left\|\boldsymbol{\phi}(\cdot)\right\|_\infty^2 := \sup_{-r_2 \leq \tau \leq 0} \left\|\boldsymbol{\phi}(\tau)\right\|_2^2$ and consider the structure of (C.1)

with $t = t_0$, it follows that there exists $\lambda > 0$ such that

$$
\begin{aligned}
&\mathsf{v}(\mathbf{x}_{t_0}(\cdot)) = \mathsf{v}(\boldsymbol{\phi}(\cdot)) \leq \boldsymbol{\eta}^\top(t_0)\lambda\boldsymbol{\eta}(t_0) + \int_{-r_2}^{0} \boldsymbol{\phi}^\top(\tau)\lambda\boldsymbol{\phi}(\tau)\mathsf{d}\tau \leq \lambda \left\|\boldsymbol{\phi}(0)\right\|_2^2 + \lambda r_2 \left\|\boldsymbol{\phi}(\cdot)\right\|_\infty^2 \\
&+ \int_{-r_1}^{0} \boldsymbol{\phi}^\top(\tau) \left(\sqrt{\mathsf{F}_1^{-1}}\boldsymbol{f}_1(\tau) \otimes I_n\right)^\top \mathsf{d}\tau\lambda \int_{-r_1}^{0} \left(\sqrt{\mathsf{F}_1^{-1}}\boldsymbol{f}_1(\tau) \otimes I_n\right) \boldsymbol{\phi}(\tau)\mathsf{d}\tau \\
&+ \int_{-r_2}^{-r_1} \boldsymbol{\phi}^\top(\tau) \left(\sqrt{\mathsf{F}_2^{-1}}\boldsymbol{f}_2(\tau) \otimes I_n\right)^\top \mathsf{d}\tau\lambda \int_{-r_2}^{-r_1} \left(\sqrt{\mathsf{F}_2^{-1}}\boldsymbol{f}_2(\tau) \otimes I_n\right) \boldsymbol{\phi}(\tau)\mathsf{d}\tau \\
&\leq (\lambda + \lambda r_2) \left\|\boldsymbol{\phi}(\cdot)\right\|_\infty^2 + \lambda \int_{-r_2}^{0} \boldsymbol{\phi}^\top(\tau)\boldsymbol{\phi}(\tau)\mathsf{d}\tau \leq (\lambda + 2\lambda r_2) \left\|\boldsymbol{\phi}(\cdot)\right\|_\infty^2
\end{aligned}
\tag{C.13}
$$

for any $\boldsymbol{\phi}(\cdot) \in \mathbb{C}\left([-r_2, 0] ⨾ \mathbb{R}^n\right)$ in (19), where (C.13) is derived via the property of quadratic forms: $\forall X \in \mathbb{S}^n, \exists\lambda > 0 : \forall \mathbf{x} \in \mathbb{R}^n \setminus \{\mathbf{0}\}, \mathbf{x}^\top \left(\lambda I_n - X\right)\mathbf{x} > 0$ together with the application of (B.2) with $\varpi(\tau) = 1$ and appropriate $\mathsf{f}(\tau)$. Consequently, the result in (C.13) shows that one can construct an upper bound of (C.1) which satisfies (29) with a $\epsilon_2 > 0$.

Now applying (B.2) to (C.1) twice with $\varpi(\tau) = 1$ and $\mathsf{f}(\tau) = \sqrt{\mathsf{F}_1^{-1}}\boldsymbol{f}_1(\tau)$, $\mathsf{f}(\tau) = \sqrt{\mathsf{F}_2^{-1}}\boldsymbol{f}_2(\tau)$ produces

$$
\begin{aligned}
&\int_{-r_1}^{0} \boldsymbol{x}^\top(t+\tau)Q_1\boldsymbol{x}(t+\tau)\mathsf{d}\tau \geq [*]\left(I_{d_1} \otimes Q_1\right) \int_{-r_1}^{0} \left(\sqrt{\mathsf{F}_1^{-1}}\boldsymbol{f}_1(\tau) \otimes I_n\right) \boldsymbol{x}(t+\tau)\mathsf{d}\tau \\
&\int_{-r_2}^{-r_1} \boldsymbol{x}^\top(t+\tau)Q_2\boldsymbol{x}(t+\tau)\mathsf{d}\tau \geq [*]\left(I_{d_2} \otimes Q_2\right) \int_{-r_2}^{-r_1} \left(\sqrt{\mathsf{F}_2^{-1}}\boldsymbol{f}_2(\tau) \otimes I_n\right) \boldsymbol{x}(t+\tau)\mathsf{d}\tau
\end{aligned}
\tag{C.14}
$$

provided that (34) holds. Moreover, by utilizing (C.14) to (C.1) with (34) and (C.13), it is clear to see that the existence of the feasible solutions of (33) and (34) infer that (C.1) satisfies (29) with some $\epsilon_1; \epsilon_2 > 0$.

In conclusion, we have shown that there exists a functional (C.1) and $\epsilon_1; \epsilon_2 > 0$ satisfying the dissipative condition in (31), and the stability criteria in (29)–(30) if the conditions in (33)–(35) are feasible for some matrices. As a result, it shows that the existence of the feasible solutions of (33)–(35) infers that the trivial solution of the closed-loop system in (19) with $\boldsymbol{w}(t) \equiv \mathbf{0}_q$ is uniformly asymptotically stable in $\mathbb{C}([-r, 0] ⨾ \mathbb{R}^n)$, and the system in (19) with (32) is dissipative.

Now consider the situation of $r_1 = r_2$ where the delay of the system in (19) is of constant values. It is not difficult to show that the corresponding synthesis condition constructed via the functional in (C.1), following the procedures (C.1)–(C.14) with $r_1 = r_2$, can be obtained by choosing $d_2 = \delta_2 = 0$ in (33)–(35) with $Q_2 = R_2 = Y = \mathsf{O}_n$. Similarly, the corresponding synthesis condition for $r_1 = 0$; $r_2 > 0$ can be obtained by choosing $d_1 = \delta_1 = 0$ in (33)–(35) with $Q_1 = R_1 = \mathsf{O}_n$. Note that the use of $\mathbb{1}, \widehat{\mathbb{1}}$ in (25) and (38), and $\widehat{\mathsf{O}}$ in (24) allows (33)–(35) to cover the corresponding synthesis conditions for the cases of $r_1 = r_2$ and $r_1 = 0$; $r_2 > 0$, without introducing redundant matrices or matrices with ill-posed dimensions.